\documentclass{ab}

\usepackage{graphicx}
\usepackage{latexsym}

\def\degr{\hbox{$^\circ$}}

\newcommand{\km}{\,\mbox{km/s}}

\def\pak{$PA_{kin}$\,}

\begin{document}

\title{Warped polar ring in the Arp\,212 galaxy}

\author{A.V.~Moiseev}
\institute{Special Astrophysical Observatory Russian Academy of Sciences,
Nizhnij Arkhyz, 369167, Russia}

\date{April 9, 2008/Revised: May 5, 2008}
\offprints{A.V.  Moiseev, \email{moisav@sao.ru}}

\titlerunning{Warped polar ring in the Arp\,212.}
\authorrunning{Moiseev}%

\abstract{The Fabry--Perot scanning interferometer mounted on the 6-m
telescope of the Special Astrophysical Observatory of the Russian
Academy of Sciences is used to study the distribution and
kinematics of ionized gas in the peculiar galaxy Arp\,212
(NGC\,7625, III\,Zw\,102). Two kinematically distinct
subsystems---the inner disk and outer emission filaments---are
found within the optical radius of the galaxy. The first
subsystem, at galactocentric distances  $r<3.5$~kpc, rotates in
the plane of the stellar disk. The inner part of the ionized-gas
disk ($r<1.5$--2~kpc) exactly coincides with the previously known
disk consisting of molecular gas. The second subsystem of ionized
gas is located at galactocentric distances 2--6~kpc. This
subsystem rotates in a plane tilted by a significant angle to the
stellar disk. The angle of orbital inclination in the outer disk
increases with galactocentric distance and reaches $50\degr$ at
$r\approx6$~kpc. The ionized fraction of the gaseous disk does not
show up beyond this galactocentric distance, but we believe that
the HI disk continues to warp and approaches the plane that is
polar with respect to the inner disk of the galaxy. Hence Arp\,212
can be classified as a galaxy with a polar ring (or a polar disk).
The observed kinematics of the ionized and neutral gas can be
explained assuming that the distribution of gravitational
potential in the galaxy is not spherically symmetric. Most
probably, the polar ring have formed via accretion of gas from the
dwarf satellite galaxy UGC~12549.

}

\maketitle

\section{Introduction}

Currently there is almost a consensus of opinion that gravitational interactions and mergers are among the most important factors in the evolution of galaxies, because such events change the observed structure, kinematics, star-formation history, etc. Polar-ring galaxies (PRG) are one of the consequences of those interactions. Such objects exhibit rings or disks consisting of gas, dust, and stars, which rotate in the plane that is approximately perpendicular (polar) to the disk of the main (``host'') galaxy. PRG are believed in most cases to owe their formation to galaxy mergers and accretion of the matter of the companion galaxy or gaseous filaments from the
intergalactic medium onto the host galaxy (Bournaud \& Combes \cite{bc03:Moiseev_n}; Combes (\cite{comb06:Moiseev_n}), and references therein).

PRG exhibit circular rotation in two mutually perpendicular planes thereby making it possible to study three-dimensional mass distribution in the galaxy
and determine the shape of its dark halo: oblateness, prolateness, and deviation from axial symmetry (Combes, \cite{comb06:Moiseev_n}). To do this, it is necessary to obtain sufficiently detailed data about the inner kinematics of PRG. Whitmore et al. (\cite{w90:Moiseev_n}) list 157 candidate polar-ring galaxies
selected mostly by their peculiar appearance. However, the number of ``real'' PRG, i.e., which exhibit rotation in orthogonal planes, is much smaller. Even
in the simplest case, where both the central galaxy and the ring are seen edge on, at least two longslit spectroscopic sections are needed to determine
the rotation pattern of both subsystems. A number of problems arise in the cases where the spatial orientations of the galaxy and ring are such that one of the rotation planes is seen at a moderate angle to the line-of-sight. First, in many cases it is impossible to establish the PRG nature of an object solely from its direct images. Second, a substantially greater number of spectroscopic sections with different slit orientations  are needed for a detailed kinematical study. In these cases it is better to determine the two-dimensional velocity field using optical panoramic (3D) spectroscopy or radio interferometry in lines of molecular and atomic gas.  The NGC\,2655 galaxy is a good illustration. It was suspected to have a polar ring due to the powerful dust
lane crossing the disk of the galaxy. A comparison of the velocity fields of gas and stars in the circumnuclear  region confirmed such an interpretation (Sil'chenko \& Afanasiev, \cite{silafan04:Moiseev_n}), and recent 21-cm line observations allowed the structure and kinematics of the polar ring to be analyzed (Sparke et al., \cite{sparke08:Moiseev_n}). Note that for such complex objects even an extensive set of both photometric and kinematical data may be insufficient for a definitive conclusion about the structure of subsystems in the galaxy considered, see, e.g., the case of UGC\,5600 (Shalyapina et al., \cite{labuda07:Moiseev_n}).

In this paper we analyze new observational data for the peculiar
galaxy Arp\,212. Although this object was suspected to be a
possible candidate PRG, no polar component has been found yet. In
Section~\ref{intro:Moiseev_n} we summarize the data related to
this galaxy. In Section~\ref{obs:Moiseev_n} we describe
observations and reduction of the data obtained with the 6-m
telescope of the Special Astrophysical Observatory of the Russian
Academy of Sciences. In Section~\ref{tilt:Moiseev_n} we show that
the velocity field of ionized gas contains two subsystems with
different rotation velocities. In the next
Section~\ref{2D:Moiseev_n} we construct a two-dimensional model of
the rotation of gas in the outer regions to explain the kinematics
of the polar ring. In Section~\ref{discuss:Moiseev_n} we discuss
the overall structure of the galaxy and finally in
Section~\ref{results:Moiseev_n} formulate the main results and
conclusions of the paper.

\section{Arp 212: History of investigations}

\label{intro:Moiseev_n}

Arp\,212 (NGC\,7625, III\,Zw\,102) is a peculiar galaxy. According
to the NED database, it belongs to SA(rs)a~pec type and its
optical diameter is \mbox{$D_{25}=1\farcm6$}. Most of the authors
classify this galaxy as an object of early morphological type---a
lenticular or even an elliptical galaxy (see the discussion in
Brosch \& Loinger, \cite{br:Moiseev_n}). At the same time, the galaxy is rich in (both atomic and molecular) gas and contains numerous HII regions,
which are indicative of violent star formation. According to
Yasuda~et~al. (\cite{co:Moiseev_n}) and Li~et~al. (\cite{li:Moiseev_n}), the current star-formation rate lies in the interval from 2.7 to  8 $M_\odot/$year.

The high intensity of emission lines of ionized gas was one of the
main reasons why Thuan \& Martin (\cite{thuan:Moiseev_n})
classified Arp\,212 as a blue compact dwarf galaxy (BCDG),
although according to Yasuda~et~al. (\cite{co:Moiseev_n}), its size ($D_{25}=11$~kpc), mass, and luminosity ((3--8)$\times10^{10}\,M_\odot$,
$L_B$=1.3\mbox{$\times10^9\,L_\odot$),} exceed substantially the
corresponding parameters for dwarf galaxies. The tradition nevertheless persists and the object continues to appear in various catalogs and lists of  BCD galaxies (Cair\'{o}s et al., \cite{cairos1:Moiseev_n}, Cair\'{o}s et al.; \cite{cairos2:Moiseev_n}; Garc\'{i}a-Lorenzo et al.,  \cite{cairos3:Moiseev_n}.

Some researchers also doubt the classification of
Arp\,212 as an early-type galaxy.  Cair\'{o}s~et~al. (\cite{cairos1:Moiseev_n}) show that even at small galactocentric distances ($r<1$ kpc) the surface-brightness profile fits well the exponential law, which is typical of a flat disk. Therefore the above authors classify the galaxy as a
late-type object (Im). 

The most well-known feature of  Arp\,212, which shows up in all
its published optical images, is the chain of dust lanes, which
form an open ring with a radius of about $15-20''$ extending along
\mbox{$PA\approx45\degr$.} The Southwestern part of the ring
stands out by its especially strong overlapping dust lanes.
Because of this feature the galaxy not only appears in Arp's list
of peculiar galaxies (Arp, \cite{arp:Moiseev_n}), but was also marked by
Whitmore~et~al. (\cite{w90:Moiseev_n}) as an object ``related to
polar-ring galaxies''.

Another interesting structural feature was found on images taken
in the H$\alpha$ emission line (Cair\'{o}s~et~al., \cite{cairos2:Moiseev_n}). Most of
the HII emission concentrates in the central part of the galaxy.
At the same time, some of the HII regions are located in
curvilinear segments extending out to 4~kpc ($35-45''$) from the
center. Cair\'{o}s~et~al. (\cite{cairos2:Moiseev_n}) believe that
these features resemble the tidal tails in the well-known Antennae
galaxy (VV 245), i.e., they are due to recent interaction.

The kinematics of the interstellar medium in  Arp\,212 was studied
repeatedly by many authors. Long-slit spectroscopy along the major
axis of the galaxy revealed a significant gradient of
line-of-sight velocity of ionized gas (Demoulin, \cite{demoulin:Moiseev_n}),
the velocity of rotation amounts up to 155~km/s. Measurements of
line-of-sight velocity of molecular gas (Yasuda et al.,   \cite{co:Moiseev_n}) agree fairly well with the model of circular rotation of the gaseous
disk. At the same time, appreciable deviations from circular
rotation were found in several regions in the vicinity of dust
lanes---possibly due to impacts of gaseous clouds onto the disk of
the galaxy. The rather low spatial resolution of these radio
observations ($beam=15''$) made it impossible to unambiguously
interpret the observed pattern of gas motions. The CO-line
velocity field constructed by Li~et~al. (\cite{li:Moiseev_n}) with a
twice higher resolution showed that the inner part of the
molecular disk rotates circularly in the plane of the stellar
disk. The above authors failed to measure the velocities of
molecular gas in the region of dust lanes, but they report the HI
velocity field and found the diameter of the HI disk to be four
times greater than the optical diameter $D_{25}$. Note that
outside the stellar disk gas rotates in the same plane, but in the
opposite direction. The HI kinematics inside the optical disk is
rather intricate with line-of-sight velocity contours turning
almost by  $90\degr$. The above authors assumed that here we are
dealing either with a strong warp of the gaseous disk, or a tilted
gaseous ring. Unfortunately, the low spatial resolution
($beam=51''$) of their observations made it impossible to study
the pattern of HI motions inside the optical disk.

There is no doubt that the galaxy bears signs of recent
interaction, however, so far there are no clear answers to the
following questions: what is the object the galaxy interacts with;
how do the outer emission regions move and whether they belong to
the galaxy at all, etc.  The answers to these questions can be
found by studying the velocity field of ionized gas in the
``transition region'' between the inner molecular disk and the
outer HI structure. In 2003   Arp212 was observed with the
scanning Fabry--Perot interferometer (FPI) attached to the 6-m
telescope of the Special Astrophysical Observatory of the Russian
Academy of Sciences (SAO RAS) at the request of
Mu\~{n}oz-Tu\~{n}\'{o}n (IAC, Spain) and her coathors. The primary
aim of the project was to analyze the effect of intense star
formation on the interstellar medium and we published the first
results in our earlier paper Mart\'{i}nez-Delgado (\cite{delgado:Moiseev_n}). However, even a preliminary analysis of the velocity field allowed us to
identify a second, independently rotating subsystem of ionized gas
in  Arp\,212, which is indicative of a warped (and, possibly,
polar) disk. The aim of this paper, which is based on archive
observational data of the SAO RAS, is to study the kinematics of
ionized gas in Arp~212.

In accordance with Li~et~al. (\cite{li:Moiseev_n}), we assume that
the distance to the galaxy is equal to  $23.5$~Mpc, which
corresponds to an image scale of 115 pc/$1''$.

\section{Observations and data reduction}
\label{obs:Moiseev_n}

\subsection{Panoramic spectroscopy}

The Arp\,212 galaxy was observed on November 30, 2003 with the
scanning Fabry--Perot interferometer mounted inside SCORPIO focal
reducer (Afanasiev \& Moiseev, (\cite{afanas05:Moiseev_n}) located in the
primary focus of the 6-m telescope of the SAO RAS. The desired
spectral interval in the neighborhood of the H$\alpha$ line was
cut using a narrow-band filter with a width of
\mbox{$FWHM=21$\AA}. The width of the free spectral interval
between the neighboring orders of interference was equal to
13\,\AA\, (about 600\km). The resolution of the interferometer
($FWHM$ of the instrumental profile) was $0.8$\AA\, ($35\km$) for
a  0.36\AA/channel scale. The detector was a $2048\times2048$ EEV
42-40 CCD operating in the $4\times4$ pixel instrumental averaging
mode in order to reduce the readout time. The resulting image
scale and field of view were 0\farcs7/pixel and 6\farcm1$\times$6\farcm1,
respectively.

We took a total of 36 successive interferograms of the object with
different gaps between the plates of the Fabry--Perot
interferometer. The total exposure was 6480~s and seeing varied
from  1\farcs3 to 1\farcs8. To remove ghost images that appear in
the plates of the Fabry--Perot interferometer employed, we
observed the object successively in two fields   turned in the position angle. We removed the ghost images
using the algorithms described by Moiseev \& Egorov (\cite{mois08:Moiseev_n}). We reduced observational data using an IDL software package (Moiseev, \cite{mois02:Moiseev_n}; Moiseev \& Egorov~\cite{mois08:Moiseev_n}). After primary reduction, night-sky line subtraction, and wavelength calibration the
observations were reduced to the form of a data cube where each
pixel in the $512\times512$ field of view contains a 36-channel
spectrum.

The final angular resolution corresponds to the  1\farcs9 seeing.
To increase the signal-to-noise ratio in low surface brightness
regions, we smoothed the data cube with a two-dimensional Gaussian
with the $FWHM$ equivalent to this seeing. With the above
procedure applied the spatial resolution became equal to 2\farcs7.
We analyzed the data with both variants of spatial resolution  and
the resulting conclusions are mutually consistent. Below we give
only the results obtained by analyzing the smoothed cube, because
this a procedure yields somewhat more accurate kinematical
parameters compared to those inferred from unsmoothed data.

We fitted the  H$\alpha$ emission profiles to the Voigt function,
which in most cases describes the observed contour fairly well
(see a discussion in Moiseev \& Egorov, \cite{mois08:Moiseev_n}). The profile fitting results were used  to construct the two-dimensional
field of line-of-sight velocities of ionized gas, the map of
velocity dispersion, and images of the galaxy in the H$\alpha$
line and in the neighboring continuum. We estimate the accuracy of
line-of-sight velocity measurements by measured signal-to-noise
ratios using the relations given in Fig.~5 of Moiseev \& Egorov (\cite{mois08:Moiseev_n}). We found this accuracy to be
\mbox{1--6\km} in bright HII regions and amount to
\mbox{10--20\km} in regions with minimal surface brightness, where
the emission line shows up at the signal-to-noise level of
$S/N\approx5$.

\subsection{Deep Image}

\begin{figure*}[tbp]
\centerline{\includegraphics[width=16 cm]{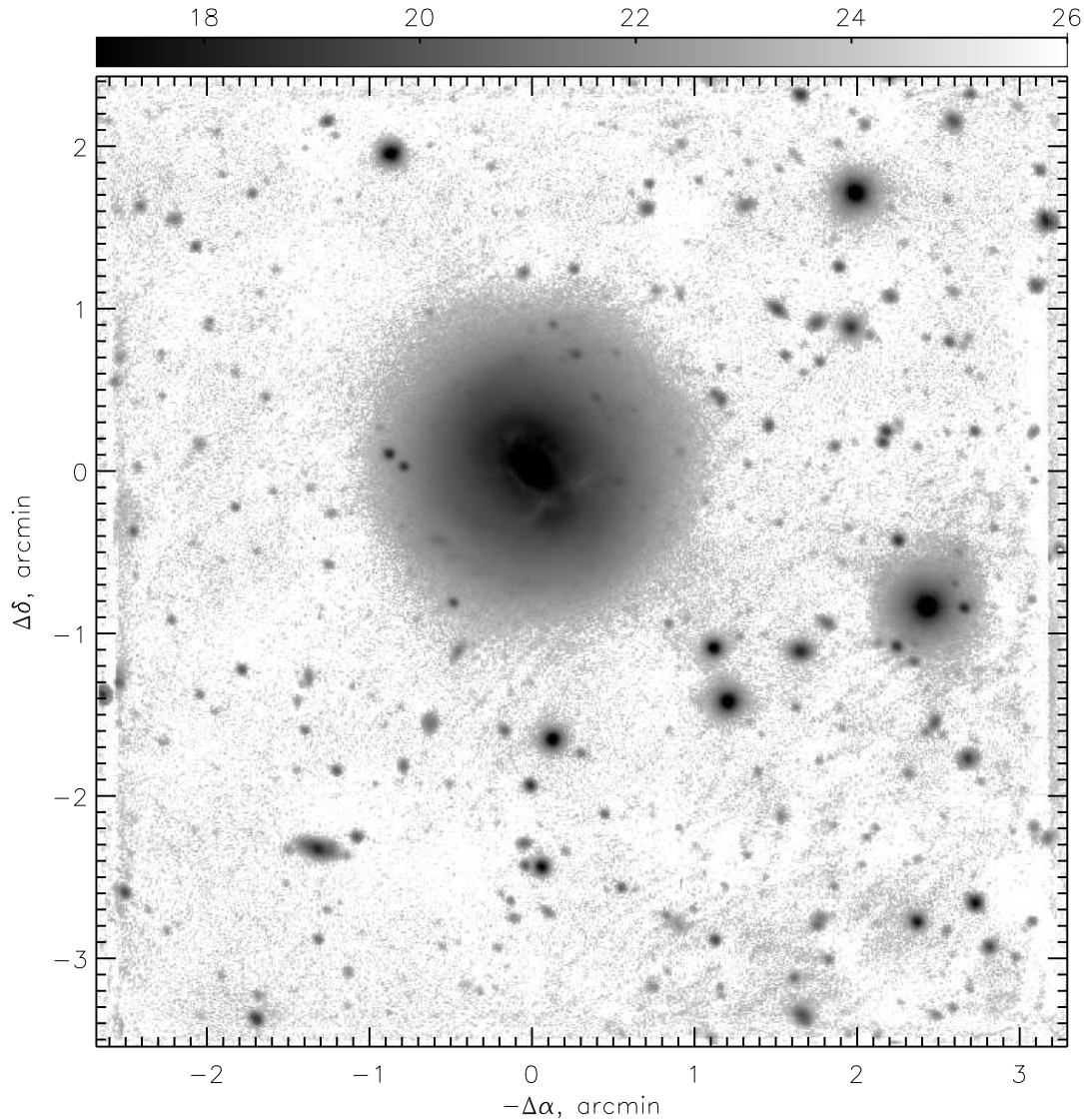}}
\caption{Deep R$_c$-band image of  Arp~212  in magnitude scale.}
\label{fig_image:Moiseev_n}
\end{figure*}

We took a direct image of Arp\,212 on February 5, 2008 with the
6-m telescope of the SAO RAS using the same instrument (SCORPIO
focal reducer). The total exposure in the R$_c$ filter of the
Johnson--Cousins system was equal to  1320~s, the seeing was
1\farcs9, the image scale 0\farcs35/pixel and the field of view
\farcm1$\times$6\farcm1. We  calibrated the image in to magnitudes using the
surface-brightness profile obtained by
Cair\'{o}s~et~al. (\cite{cairos1:Moiseev_n}) in the same filter.
Although observations were made under nonphotometric atmospheric
conditions at zenith distances  $z$=61--71$\degr$, we managed to
obtain so far deepest available image of the galaxy. The limiting
detection magnitude at the $1\sigma$ level is equal to
$25.2-25.5^{m}/\Box''$. The visible radially averaged
surface-brightness profile extends one and a half times farther
from the center of the galaxy compared to the profile obtained by
Cair\'{o}s~et~al. (\cite{cairos1:Moiseev_n}) and reaches a surface
brightness level of $26^{m}/\Box''$. The image was taken in order
to search for eventual low surface brightness tidal features in
the vicinity of the galaxy. However, it is evident from
Fig.~\ref{fig_image:Moiseev_n} that no external filaments show up
in the field of view.

We decomposed the surface-brightness profile into standard
components: exponential disk (with a radial scale $h=$10\farcs3,
which is close to the result of 
Cair\'{o}s~et~al. (\cite{cairos1:Moiseev_n}), and a central surface
brightness of  $\mu_0=17.4^m$) and S\'ersic's bulge with parameters
$n=1$, $r_e=$5\farcs2, and $\mu_e=17.7^m$. The central surface
brightness of the disk far exceeds the average central surface
brightness for disk galaxies, as it is typical for interacting
systems (Reshetnikov et al.,\cite{reshetnikov93:Moiseev_n}). The components luminosity ratio $B/D=0.20$ allows us to classify the galaxy as a late-type
system (Sc). Note that because the exponent in S\'ersic's law is $n
= 1$ and the inclination of the galaxy is small, we cannot
distinguish the bulge from an inner exponential disk and therefore
the contribution of the spherical component may be even smaller
and the morphological type of the galaxy may be even later.

\section{Analysis of the line-of-sight velocity field}

\label{tilt:Moiseev_n}

\subsection{Circular Rotation Model}

\begin{figure*}[tbp]
\centerline{\includegraphics[width=8. cm]{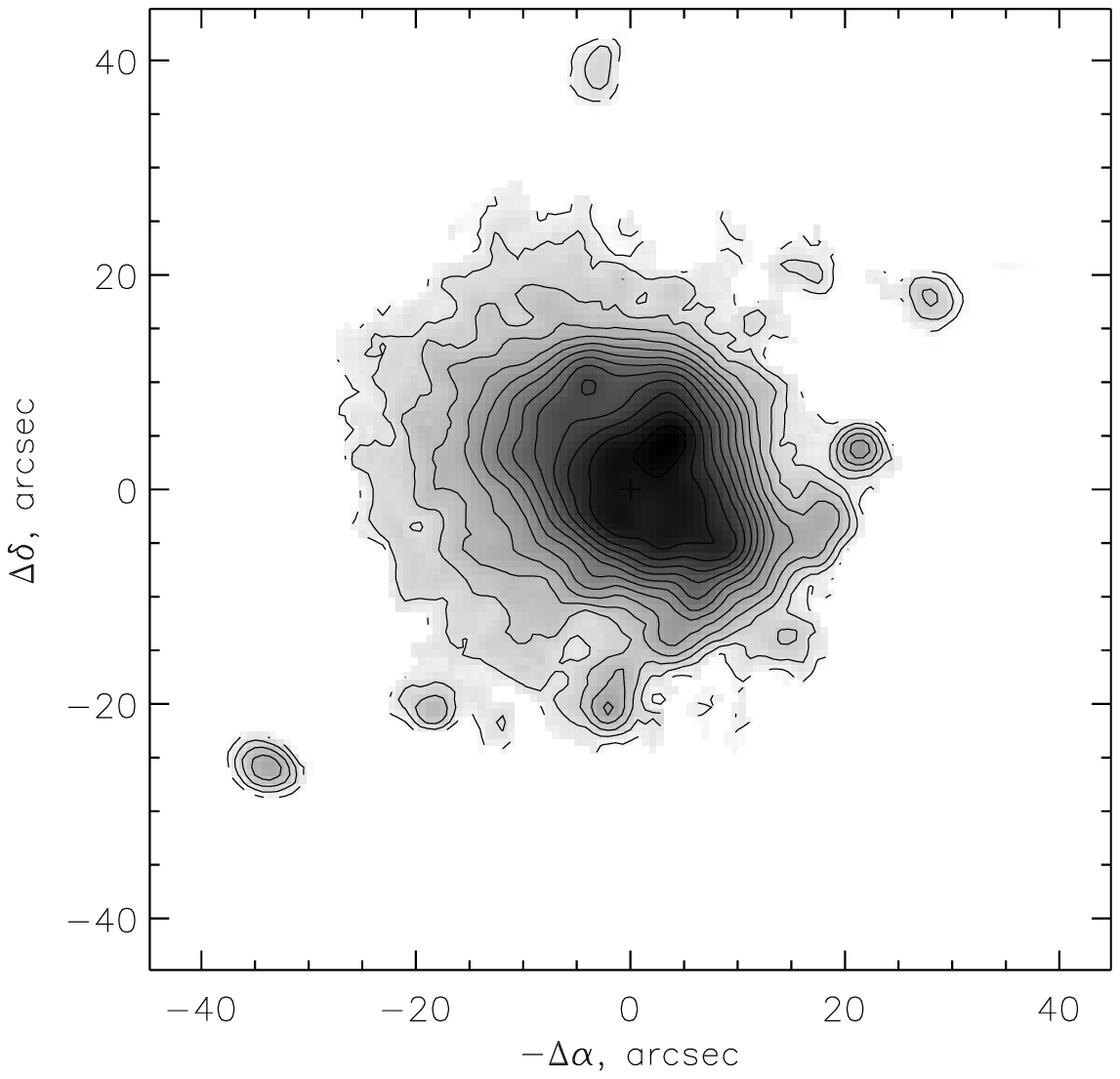}
\includegraphics[width=8.cm]{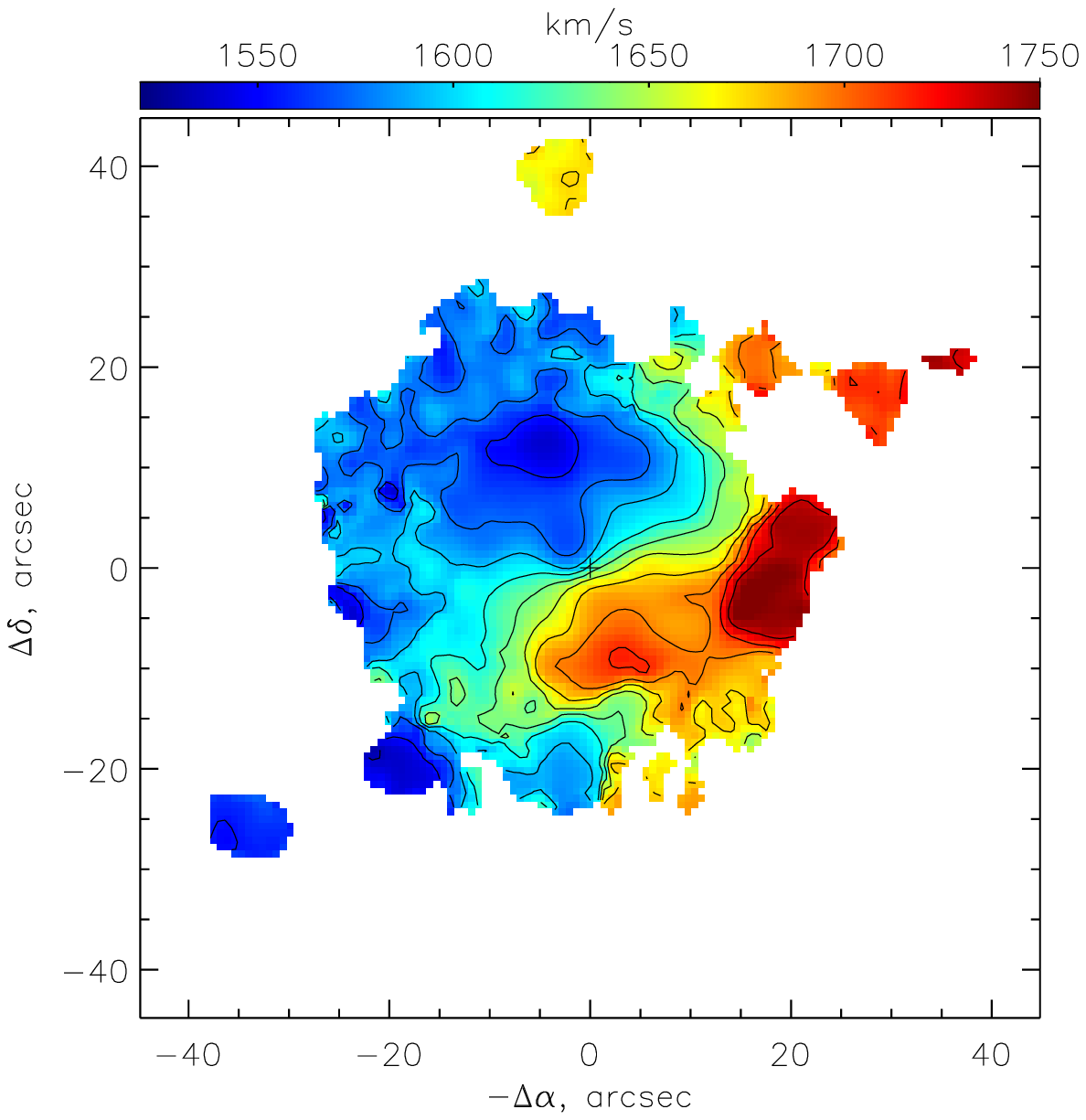}}
\caption{Results of H$\alpha$ observations of Arp~212: intensity
distribution in logarithmic scale (left panel) and line-of-sight
velocity field (right panel). The coordinate origin coincides with
the center of continuum isophotes.}\label{fig_ifp:Moiseev_n}
\end{figure*}

In Fig.~\ref{fig_ifp:Moiseev_n} one can see the H$\alpha$ image of
the galaxy and the line-of-sight velocity field of ionized gas. We
confidently measured the line-of-sight velocities in almost all
outer emission regions found by Cair\'{o}s~et~al. (\cite{cairos2:Moiseev_n}). It is evident from the figure that the form of line-of-sight velocity contours corresponds to the  circular rotation in the region within
\mbox{$r\approx20''$,} i.e., where most of
the H$\alpha$ emission is concentrated. The pattern of the
line-of-sight velocity field agrees well with the results of
integral-field spectroscopy for the inner region reported by
Garc\'{i}a-Lorenzo~et~al. (\cite{cairos3:Moiseev_n}). The location
of the center of rotation as  inferred assuming that the velocity
field is symmetric agrees within the errors with the center of the
H$\alpha$ and continuum brightness contours. We analyzed the
velocity field using the method of ``tilted rings''
Begeman (\cite{beg89:Moiseev_n}). We subdivided the velocity fields into
$1''$-wide elliptical rings oriented according to the adopted
position angle of the major axis ($PA_0$) and inclination of the
disk ($i_0$). We fixed the position of the center of rotation. In
the first approximation we estimated  $PA_0$ and  $i_0$ from
photometric results. We then determined in each ring the optimum
position angle of the kinematical axis \pak, the mean velocity of
rotation $V_{rot}$ and systemic velocity $V_{sys}$. For the
justification of the method and references to the original works,
see, e.g., Moiseev \& Mustsevoi (\cite{mois00:Moiseev_n}) and Moiseev et al. (\cite{mois04:Moiseev_n}).

\begin{figure}[tbp]
\centerline{\includegraphics[width=8. cm]{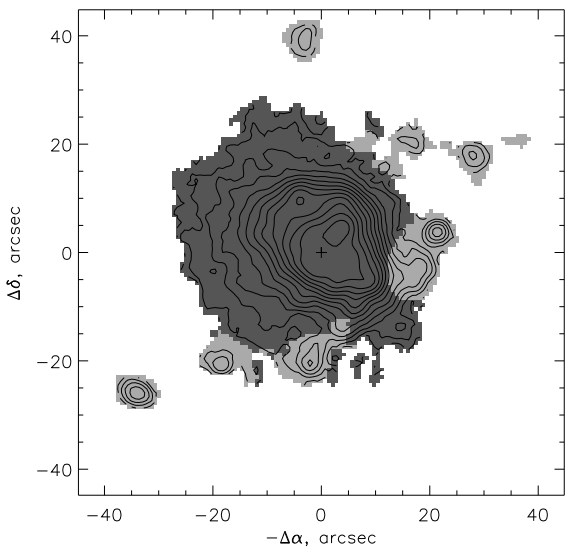}}
\caption{The subdivision of galaxy into domain
of the inner disk (dark shaded) and outer filaments (light shaded)
with H$\alpha$-line isophotes superimposed.}
\label{fig_regions:Moiseev_n}
\end{figure}

Our preliminary analysis revealed two groups of domains in the
velocity field: domains with velocities close to those implied by
the circular-rotation model at the given galactocentric radius
$r$, and domains with velocities that appreciably differ from the
circular-rotation velocities by more than a certain threshold
value ($v_{lim}$). We constructed several variants of the
circular-rotation model choosing  $v_{lim}$ so as to make data
points belonging to each velocity group concentrate in spatially
connected domains in the sky plane. We found that the best result
is achieved with $v_{lim}\approx36$ km/s. In this case two types
of domains show up conspicuously in
Fig.~\ref{fig_regions:Moiseev_n}: the inner disk with
close-to-circular rotation and isolated kinematically distinct
domains. The latter include all HII regions at galactocentric
distances greater than $20''$ and several HII regions projected
onto the Southwestern edge of the disk of the galaxy.

Table~\ref{tab:Moiseev_n} lists the orientation parameters for the
inner disk of the galaxy inferred from the velocity field assuming
that $PA$, $i$, and $V_{sys}$ remain constant along the radius
(the model of purely circular rotation in a fixed plane). These
parameters agree within the $3\sigma$ with those found by
Li~et~al. (\cite{li:Moiseev_n}) for the velocity field of the
molecular gas disk at \mbox{$r\le15''$.}

\subsection{Quasi-circular rotation model}

\subsubsection{Inner disk}

\begin{figure}[tbp]
\centerline{\includegraphics[width=8. cm]{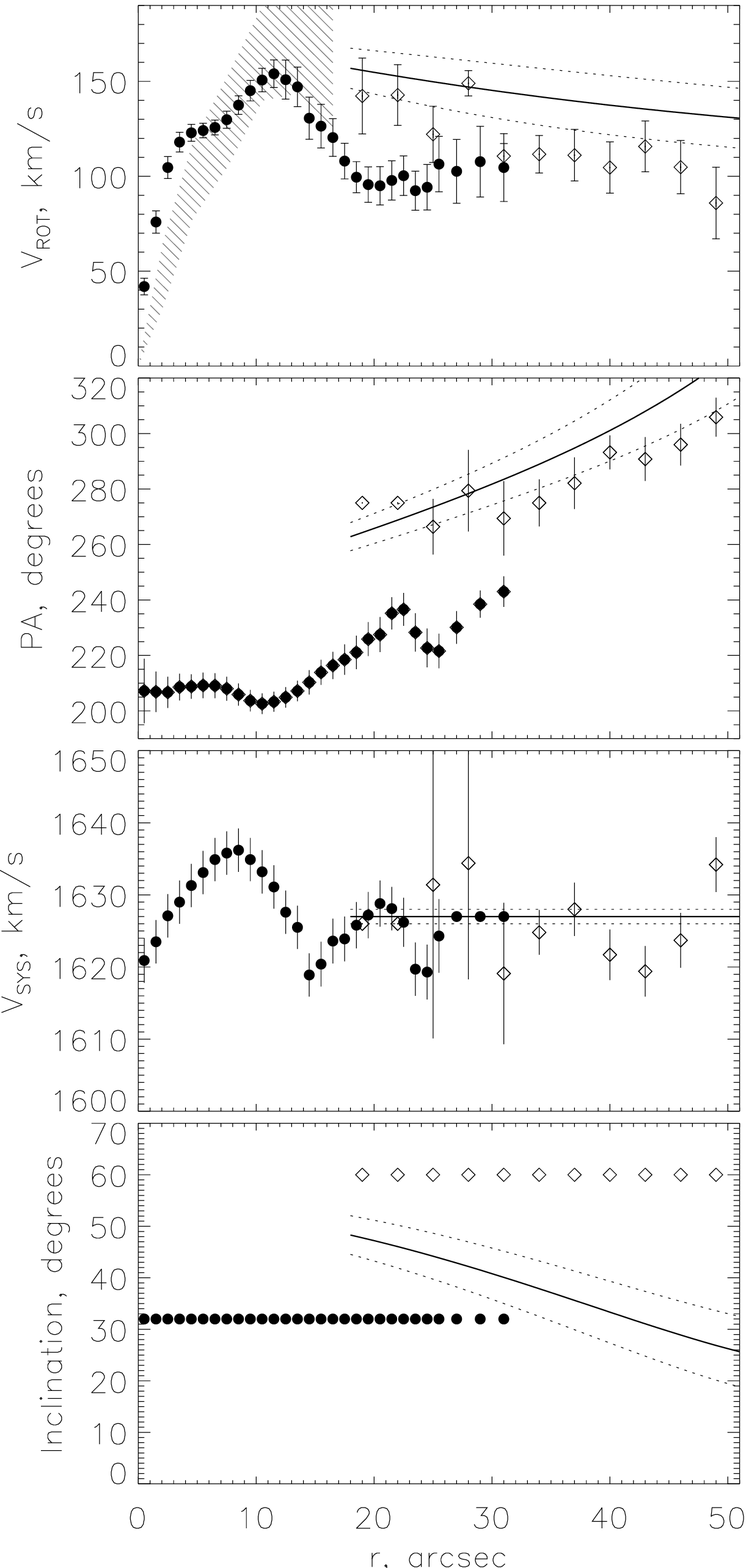}}
\caption{Radial variations  of kinematic parameters characterizing the velocity field of ionized gas (from top to bottom): circular-rotation velocity; position angle, and inclination. The filled circles and open diamond signs correspond
to the inner disk and outer emission regions, respectively. The
hatched domain on the upper plot reflects the rotation velocity of
molecular gas (Li et al., \cite{li:Moiseev_n}) with the measurement errors
taken into account. The solid lines correspond to the kinematical
parameters computed in terms of a two-dimensional warped-disk
model (see Section~\ref{2D:Moiseev_n}). The dashed lines indicates
the $3\sigma$ interval computed from errors of model parameters
from Table~\ref{tab2:Moiseev_n}.
 }\label{fig_rc:Moiseev_n}
\end{figure}

We refined the above model using the quasi-circular approximation,
which allows noncircular motions, disk warp, etc. to be taken into
account. We assumed that gas rotates in circular orbits whose
$V_{sys}$ and  \pak vary with $r$. Results of
the analysis are presented in Fig.~\ref{fig_rc:Moiseev_n}.
Systemic velocity does not deviate from the mean value determined
above by more than \mbox{$\pm10\km$}. The rotation curve of the
inner part of the disk is close to the molecular-gas data,
adjusted to our adopted value for inclination  $i_0$. However,
there are also certain systematic differences. First, our rotation
curve shows a much sharper gradient of the rotation velocity of
ionized gas, which must be due to the 2.5 times lower resolution
of radio observations of Li~et~al. (\cite{li:Moiseev_n}). Second,
the maximum rotation velocity of the molecular gas is somewhat
higher than the rotation velocity of the ionized gas, although two
velocities agree within the measurement errors. This difference
may be due to asymmetric drift, i.e., to different velocity
dispersions of the components studied. Indeed the velocity
dispersion of ionized gas outside \mbox{$r>10''$} exceeds $50
\km$, and such a behavior is indicative of the additional increase
of chaotic motions caused by a powerful burst of star
formation (Mart\'{i}nez-Delgado, \cite{delgado:Moiseev_n}).

A systematic increase of \pak can be seen in the  interval  $r$=13--$30''$ and the total amplitude of its variations exceeds $30\degr$. This phenomenon can be most naturally explained by the warp of the outer disk regions.
Unfortunately, because of the rather small $i_0$ the method
employed is unstable with respect to parameter $i(r)$---the tilt
of the disk with respect to the sky plane inside each narrow ring.
Therefore the adopted model cannot be used to determine the
variation of the tilt of the warped part of the disk with respect
to the line-of-sight.

\subsubsection{Outer regions}

\begin{figure*}[tbp]
\centerline{\includegraphics[width=8. cm]{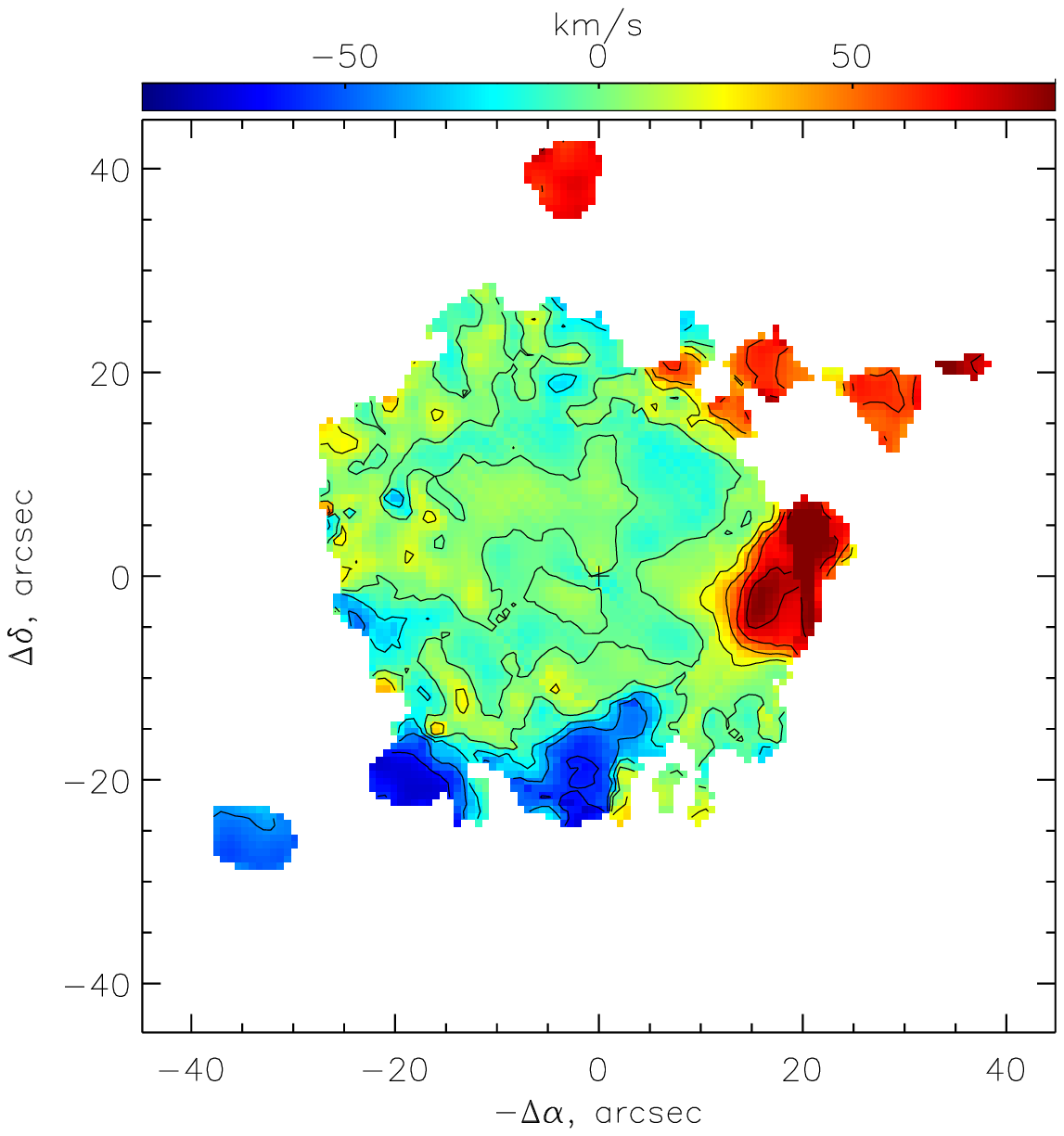}
\includegraphics[width=8.cm]{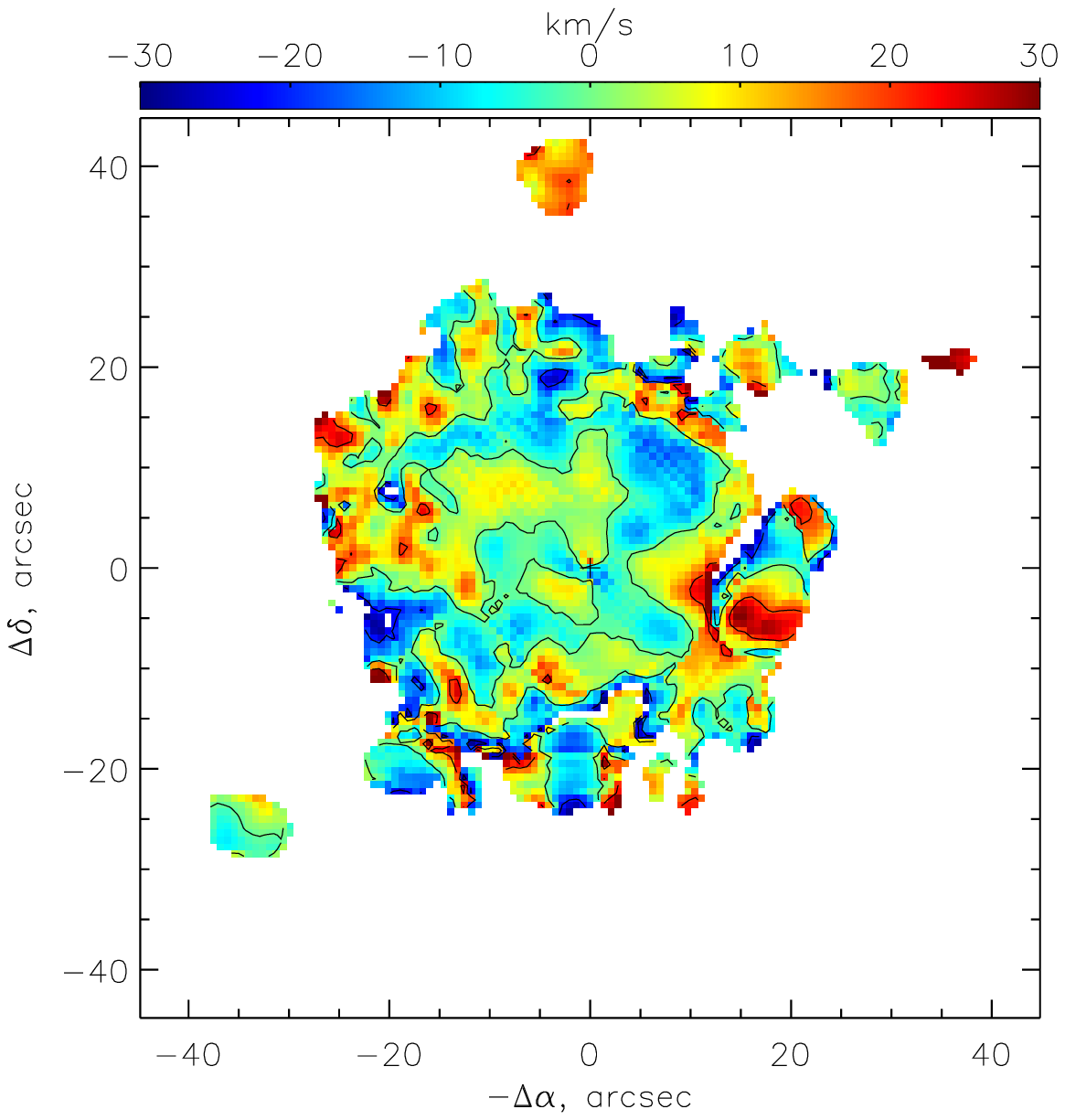}}
\caption{Maps of the residual line-of-sight velocities (observed
minus model): cases of a circular rotation model (left panel) and
separate accounting for the rotation of inner and outer regions
(right panel). }\label{fig_res:Moiseev_n}
\end{figure*}

On the left panel of Fig.~\ref{fig_res:Moiseev_n} we present the
map of residual velocities (observed minus model). Here we
extrapolated the model up to outer filaments. Residual velocities
in the inner disk mostly do not exceed $\pm20\km$, whereas in the
outer HII regions they are several times higher and exhibit
well-defined systematic behavior varying smoothly from  $+100\km$
in the Northwest to $-75\km$ in the Southeast. The pattern of the
velocities of outer filaments resembles the rotation of an
inclined disk with orientation parameters differ appreciably from
those of the inner region.

We determined the parameters of the rotation model for outer
regions in terms of a purely circular-rotation model like we did
it above for the inner disk. We list these parameters in the last
column of Table~\ref{tab:Moiseev_n}. The systemic velocities of
both kinematic components coincide, whereas their $PA_0$ and $i_0$
differ substantially. The center of rotation of outer regions is
located  $1''$ from the center of the inner disk, i.e., both
centers coincide.

The diamond signs in Fig.~\ref{fig_rc:Moiseev_n} show the results
of our analysis of the velocity field of filaments in  terms of
quasi-circular approximation. We performed our analysis assuming
that angle $i$ remains constant along the radius. We again failed
to find the variations of $i(r)$ due to rather small number of
data points inside each narrow ring. The position angle of the
line of nodes for circular orbits of filaments varies with
galatcocentric radius so that in the outer regions the turn of
$PA$ with respect to the inner disk amounts to  $90\degr$ and
exceeds appreciably the $PA_0$ value listed in
Table~\ref{tab:Moiseev_n}. The variations of \pak about its mean
value are usually interpreted either as signature of noncircular
motions in the disk plane or as manifestations of the orientation
(the warp) of the disk with respect to the line-of-sight. We
believe the former hypothesis to be unlikely, because a turn of
\pak by several tens of degrees would require substantial radial
and azimuthal motions (exceeding $50\km$), which must be
comparable to the velocity of circular motion. Such a situation is
sometimes observed in barred galaxies, but it is by no means
typical of outer disk regions, since it require a substantial
asymmetry in the distribution of gravitational potential (spiral
density waves and a bar). We believe the second interpretation
(the turn of the plane of circular orbits) to be most realistic,
since the warp of the outer gaseous disk is a common phenomenon,
especially in interacting galaxies. Below we consider the possible
orientations of the warped disk with respect to the plane of the
galaxy.

\begin{table}[tbp]
\captionstyle{flushleft}
\caption{Average kinematical parameters of gaseous subsystems.} \label{tab:Moiseev_n}\bigskip
\begin{tabular}{lrr}\hline\hline
Parameter& Inner disk &  External parameters\\
         & (at $r=0-20''$) & (at $r=25-50''$) \\
\hline
$PA_0\,(\degr)$ & $211\pm6$ & $275\pm15$\\
$i_0\,(\degr)$ & $32\pm8$ & $60\pm10$\\
$V_{sys}\,(\km)$ & $1627\pm5$ &  $1626\pm6$\\
\hline
\end{tabular}
\end{table}

\begin{table}[tbp]
\captionstyle{flushleft}
\caption{Parameters of the two-dimensional model.} \label{tab2:Moiseev_n}
\bigskip
\begin{tabular}{lr}\hline\hline
Parameter& Value\\
         &     \\
\hline
$\delta_0\,(\degr)$    & $29\pm3$\\
$\delta_1\,(\degr/'')$ &$0.40\pm0.06$\\
$\gamma_0\,(\degr)$    &$51\pm5$\\
$\gamma_1\,(\degr/'')$& $0.94\pm0.15$\\
$V_{max}\,(\km)$        & $89\pm7$\\
$R_h\,('')$             & $-46\pm25$\\
$V_{sys}\,(\km)$        & $1627\pm1$\\
\end{tabular}
\end{table}

\subsection{Spatial Orientation of Orbits}

The significant differences between the orientation parameters of
the inner and outer gaseous subsystems (Table~\ref{tab:Moiseev_n})
indicate that the rotation plane of outer filaments is tilted
substantially to the plane of the inner disk. Angle $\delta$
between the two planes is given by the scalar product of their
directing vectors via the following formula:

$$
\cos \delta  =\pm\cos (PA_2-PA_1) \sin i_1 \sin i_2  +\cos i_1 \cos i_2.
$$

Here subscripts ``1'' and ``2'' indicate the observed orientation
parameters of both disks. The sign ambiguity in the first term is
due to the fact that the position angle of the major axis,  $PA$,
and inclination $i$ as determined via kinematical analysis do not
fully characterize the position of the plane with respect to the
observer---we must know which side is the closest and which is the
farthest from us. For the ambiguity to be resolved, the formula
should be written as:

{\small
\begin{equation}
\cos \delta  =\cos (PA_2-PA_1)  \sin i_1 \sin i_2  +\cos i_1 \cos i_2,
\label{eq1:Moiseev_n}
\end{equation}
}

\noindent specifying how the inclination $i$ should be counted and
taking into account the direction of rotation of the disks. If the
vector of angular momentum is directed toward the observer, then
$i<90\degr$, if the angular momentum points to the opposite
direction, then the inclination is defined as $180\degr-i$.

The dust lanes mentioned in Section~\ref{intro:Moiseev_n} and the
dust ring are most conspicuous in the Southern and Southeastern
parts of the galaxy disk (see the image shown in
Fig.~\ref{fig_image:Moiseev_n}). The color-excess maps
(Cair\'{o}s, \cite{cairos2:Moiseev_n}) also indicate that the Southeastern part of the galaxy stands out because of its redder color, which is due
to the dust environment. This conclusion is confirmed by the
Balmer-decrement maps reported by Garc\'{i}a-Lorenzo~et~al. (\cite{cairos3:Moiseev_n}). The observed
asymmetry of the distribution of extinction implies that the
Southeastern part of the stellar disk of Arp\,212 is the closest
to us, and the pattern of the velocity fields implies that the
vector of angular momentum is directed toward us.

It is evident from Fig.~\ref{fig_regions:Moiseev_n} that most of the external emission regions concentrate in the Southeast part of the galaxy, whereas there are no anomalous (compared to HII regions of the
inner disk) HII regions in the  position-angle interval \mbox{$PA\approx0$--$130\degr$}. It is safe to assume
that this part of the outer disk is hidden from the observer by the stellar disk of the galaxy. In this case the Southeastern part of the orbits of emission filaments is the closest to us and the angular-momentum vector is
directed toward the observer.

In view of the above discussion of the orientation of the disk
with respect to the observer we use Eq.~(\ref{eq1:Moiseev_n}) to
compute the angle between the two planes for $i_1=32\degr$,
$i_2=60\degr$, $PA_1=211\degr$ assuming that $PA_2$ varies with
radius as shown in Fig.~\ref{fig_rc:Moiseev_n}. Namely,
\mbox{$\delta=(45\pm11)\degr$} for the inner part of the warped
disk (\mbox{$r=20''$}), whereas the angle increases with
galactocentric distance and reaches $(68\pm10)\degr$ at $r=50''$.

\section{2D model}
\label{2D:Moiseev_n}

The above investigation of the kinematics of gas performed using
the  method of ``tilted rings'' constitutes the traditional
approach toward the analysis of the velocity fields of galactic
disks. Although we managed to obtain a number of quantitative
estimates and conclusions concerning the behavior of outer gaseous
filaments, this approach has nevertheless a certain disadvantage.
Because of the insufficient sampling of the velocity field by data
points at different position angles we had to assume that the tilt
of orbits in the outer disk with respect to the line-of-sight does
not vary with radius. At the same time, we showed that this disk
is appreciably warped, so that the inclination of orbits with
respect to the inner regions varies with radius. To better analyze
the behavior of such a disk, we used the two-dimensional model of
a warped disk described by Coccato~et~al. (\cite{coccato07:Moiseev_n}). In this model the disk is subdivided into narrow rings and the orientation of $n$-th ring with respect to the main plane of the galaxy is specified by
vertical angle\footnote{For the sake of convenience, we maintain,
where possible, the designations adopted by
Coccato~et~al. (\cite{coccato07:Moiseev_n}).} $\delta_n$ and
azimuthal angle $\gamma_n$. We assume that these angles vary
linearly with galactocentric distance $r_n$:

\begin{equation}
\delta_n(r_n)=\delta_0+\delta_1\,r_n, \;\;\gamma_n(r_n)=\gamma_0+\gamma_1\,r_n.
\label{eq2:Moiseev_n}
\end{equation}

For the rotation curve of the disk we adopt the following widely used approximation (Courteau, \cite{courteau97:Moiseev_n}; Coccato, \cite{coccato07:Moiseev_n}):
\begin{equation}
V_{rot}(r_n)=\frac{2}{\pi}V_{max}\arctan\frac{r_n}{R_h}.
\label{eq3:Moiseev_n}
\end{equation}

Formulae (\ref{eq2:Moiseev_n}) and (\ref{eq3:Moiseev_n})
approximate fairly well the results obtained in terms of the model
described in the previous section.

For each ring we compute the distribution of line-of-sight
velocities in accordance with rotation
curve~(\ref{eq3:Moiseev_n}). All the necessary formulae can be
found in (Coccato, \cite{coccato07:Moiseev_n}). The contribution of each ring
to the total velocity field was assumed be the same for all rings.
Unlike Coccato~et~al. (\cite{coccato07:Moiseev_n}), we did not
simultaneously approximate the surface-brightness distribution of
the warped disk because of its ``clumpy'' appearance.

This approach fundamentally differs from the methods described in
Section~\ref{tilt:Moiseev_n}: here we construct the model for the
entire velocity field and fit it to observations. We fold the
model velocity field with the two-dimensional Gaussian with a
half-width corresponding to the spatial resolution of our data.

The $\chi^2$ minimization yields seven parameters, which fully
characterize the warped disk: $\delta_0$, $\delta_1$, $\gamma_0$,
$\gamma_1$, $V_{max}$, $R_h$, and systemic velocity $V_{sys}$. We
developed   WARPWID software package for modeling and used  MPFIT
library written by B. Markwardt\footnote{The most recent
version of MPFIT is available at {\tt
http://cow.physics.wisc.edu/\~{}craigm/idl/idl.html}} for
nonlinear minimization. As the initial approximation for the
parameter values we used the estimates obtained in terms of the
model of ``tilted rings'' described in
Section~\ref{tilt:Moiseev_n}. We list the resulting parameter
values in Table~\ref{tab2:Moiseev_n}. The quoted errors correspond
to  $3\sigma$ deviations. Multiple runs with slightly different
parameters showed that the solution obtained is stably
reproducible and depends little on the initial conditions.

\begin{figure*}[tbp]
\centerline{\includegraphics[width=8. cm]{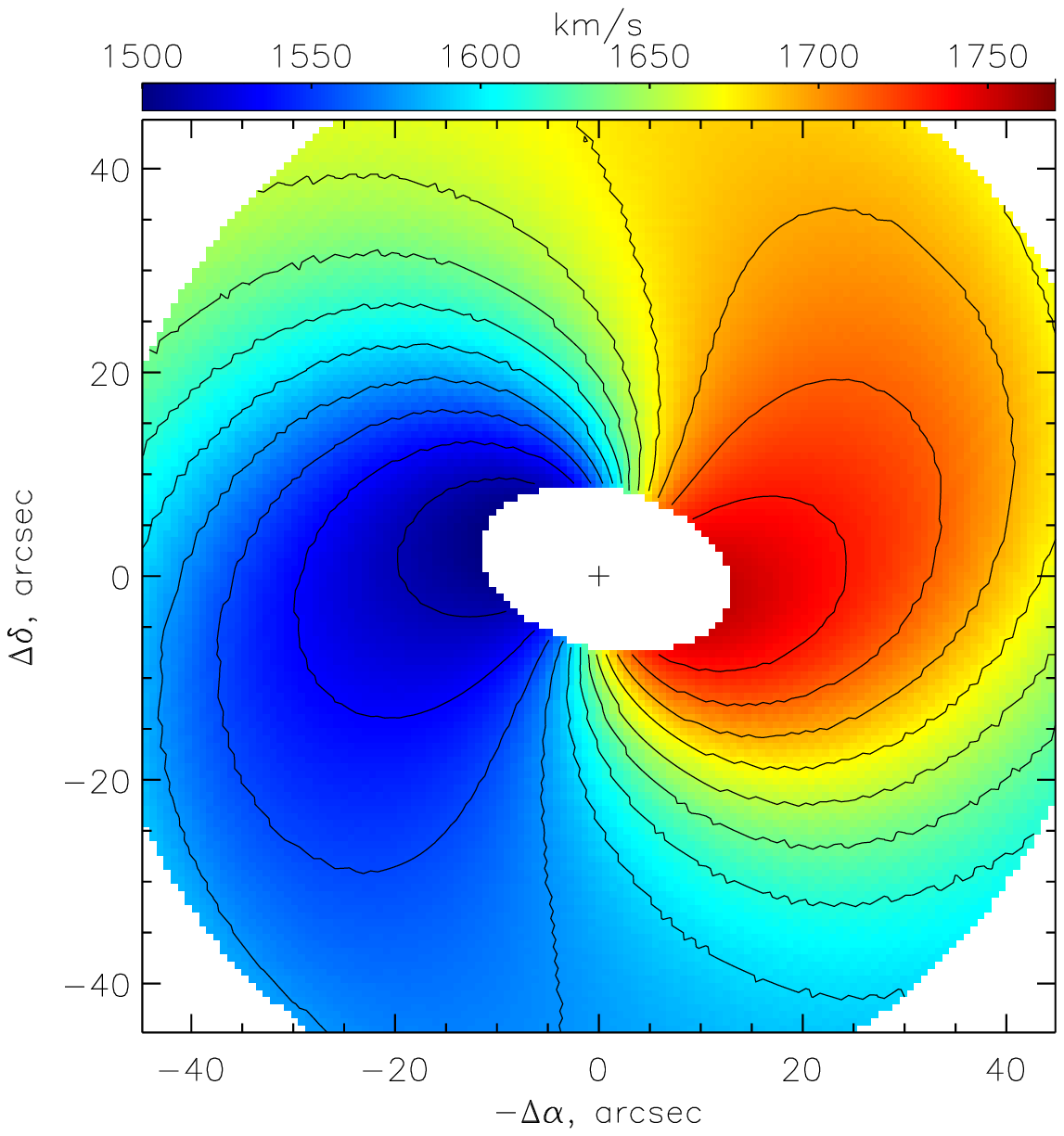}
\includegraphics[width=8.cm]{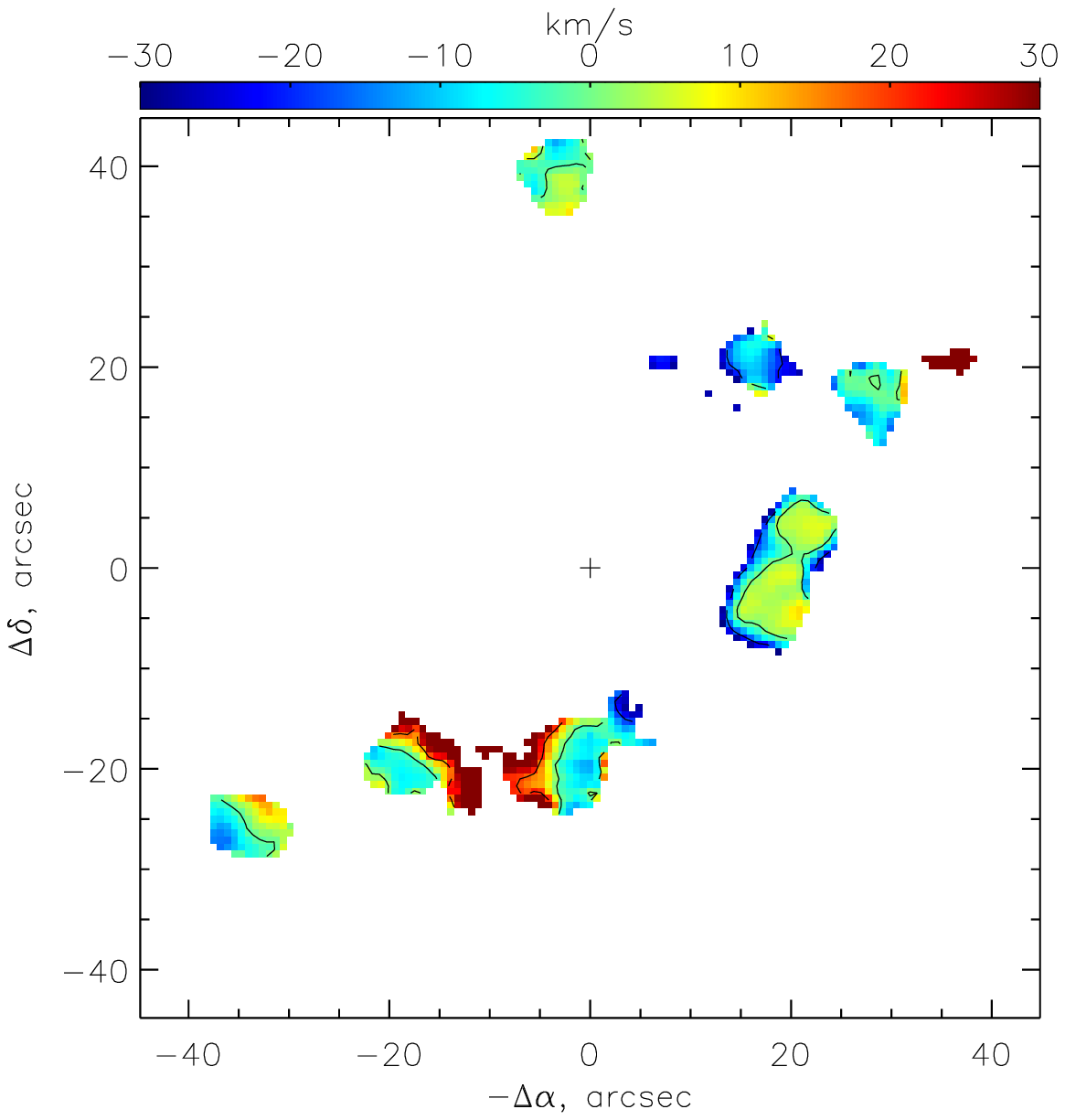}}
\caption{Velocity field in the 2D warped-disk model
(left panel) and the field of residual line-of-sight velocities in
outer filaments (right panel).}
 \label{fig_warp:Moiseev_n}
\end{figure*}

In Fig.~\ref{fig_warp:Moiseev_n} we represent the model velocity
field and the field of residual line-of-sight velocities for the
outer filaments. The deviations of observed velocities from the
corresponding model velocities are somewhat smaller than in
Fig.~\ref{fig_res:Moiseev_n}. The only exceptions are the inner
boundaries of two HII regions South of the nucleus, where residual
velocities exceed \mbox{25--30\km.} These data points most likely
correspond to the transition region where the velocities of the
inner disk and outer HII regions are projected simultaneously onto
the line-of-sight.

The solid lines in Fig.~\ref{fig_rc:Moiseev_n} reflect the radial
dependences of the kinematical parameters computed in terms of our
two-dimensional model: the rotation curve, position angle of
orbits in the warped disk, the inclination of the orbits with
respect to the sky plane, and systemic velocity (which is constant
along the radius). It is evident from the figure that  $V_{sys}$
and $PA$ agree with the results obtained in terms of the
approximation of ``tilted rings''\footnote{Note that model
dependences shown by the lines in Fig.~\ref{fig_rc:Moiseev_n}  are
not approximations of the results of computations made in terms
the ``tilted rings'' approximation. So, in each of
the models considered in Sections~\ref{tilt:Moiseev_n} and
\ref{2D:Moiseev_n} the same sky-plane point corresponds to
different galactocentric distances $r$ in 3D space, since the
orientations of orbits differs between these two models.} within the
quoted errors. The rotation velocity in the two-dimensional model
varies with radius, but it is about 30\% higher than the velocity
inferred in terms of the approximation of ``tilted rings''.
Correspondingly, the inclination of circular orbits to the sky
plane in case of the two-dimensional model differs substantially
from that adopted in Section~\ref{tilt:Moiseev_n}---it appreciably
decreases with radius. Note that the angle between the inner and
warped disks is, according to Table~\ref{tab2:Moiseev_n}, equal to
$\delta=(37\pm3)\degr$ at $r=20''$ and reaches  $(49\pm4)\degr$ at
$r=50''$. Thus in the two-dimensional model this angle also
increases with distance, albeit somewhat slowly than in case of
the `tilted rings'' approximation.

Let us now point out the main difference between the kinematical
models considered. The approximation of ``tilted rings'' requires
more than 30 parameters to describe the velocity fields of outer
emission filaments: the common inclination $i$ and three
parameters (\pak, $V_{rot}$, $V_{sys}$) for each narrow ring. The
small number of points with measured velocities inside each ring
and their nonuniform distribution in position angle make it is
impossible to measure the variations of  $i$ with 
radius and the resulting model is unstable. In the two-dimensional
model the velocity field is described only by seven parameters.
This model is much more stable, but its weak point is use of {\it
a priori} formulae for the rotation curve (\ref{eq2:Moiseev_n})
and the law of variation of orientation angles
(\ref{eq3:Moiseev_n}). These formulae were proposed based on the
results of the analysis performed using the method of ``tilted
rings'', i.e., our two-dimensional model is actually the next
approximation to the description of the kinematics of the gaseous
disk, which refines and supplements the results reported in
Section~\ref{tilt:Moiseev_n}. That is why below when discussing
the kinematics of the outer gaseous disk we refer to the results
obtained in terms of the two-dimensional model, since we consider
it to be more realistic.

\section{Discussion}

\label{discuss:Moiseev_n}

\subsection{Three-dimensional structure of the gaseous subsystem of Arp~212}

\begin{figure*}[tbp]
\centerline{\includegraphics[width=8.0cm]{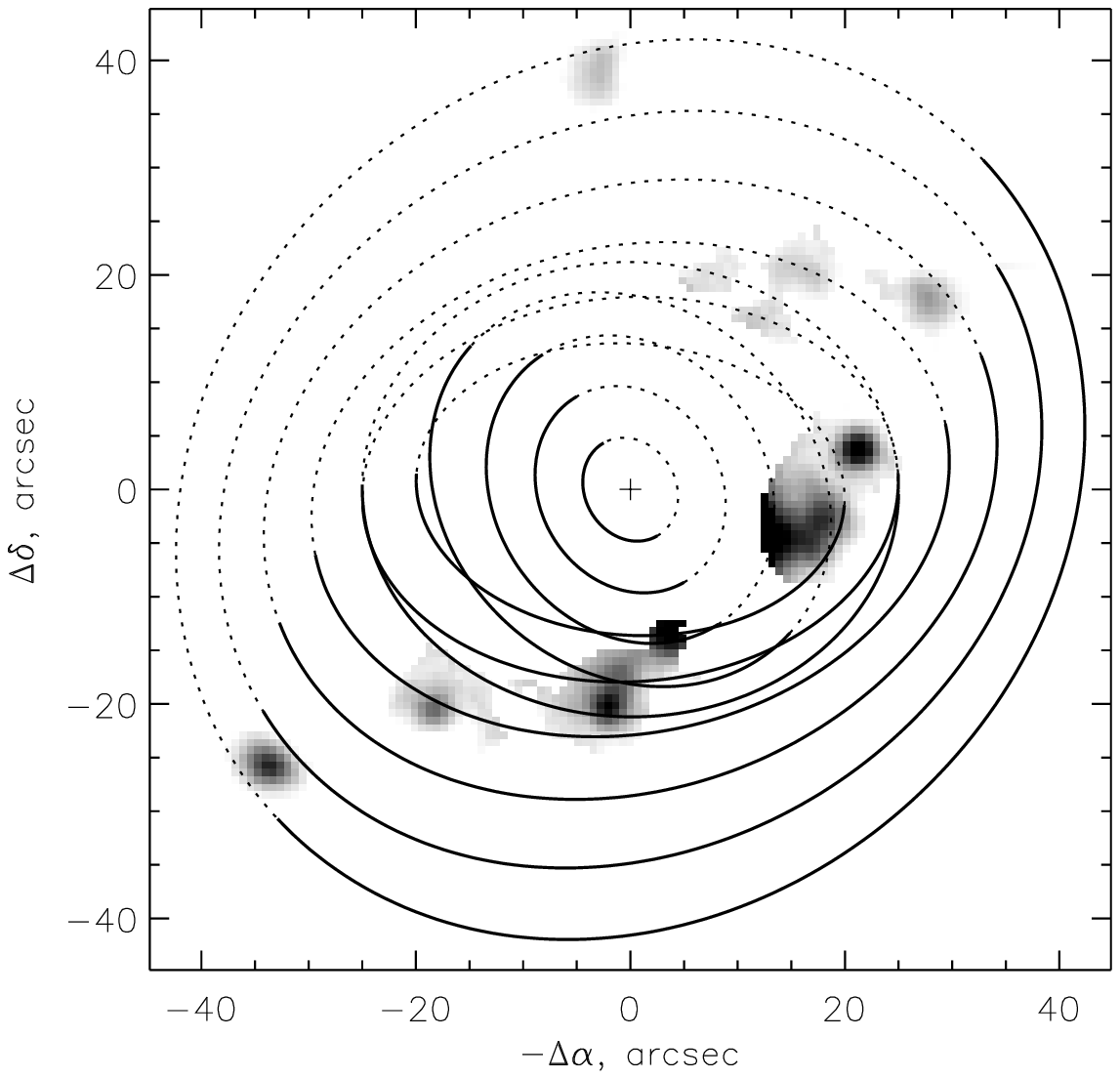}
\includegraphics[width=8.0cm]{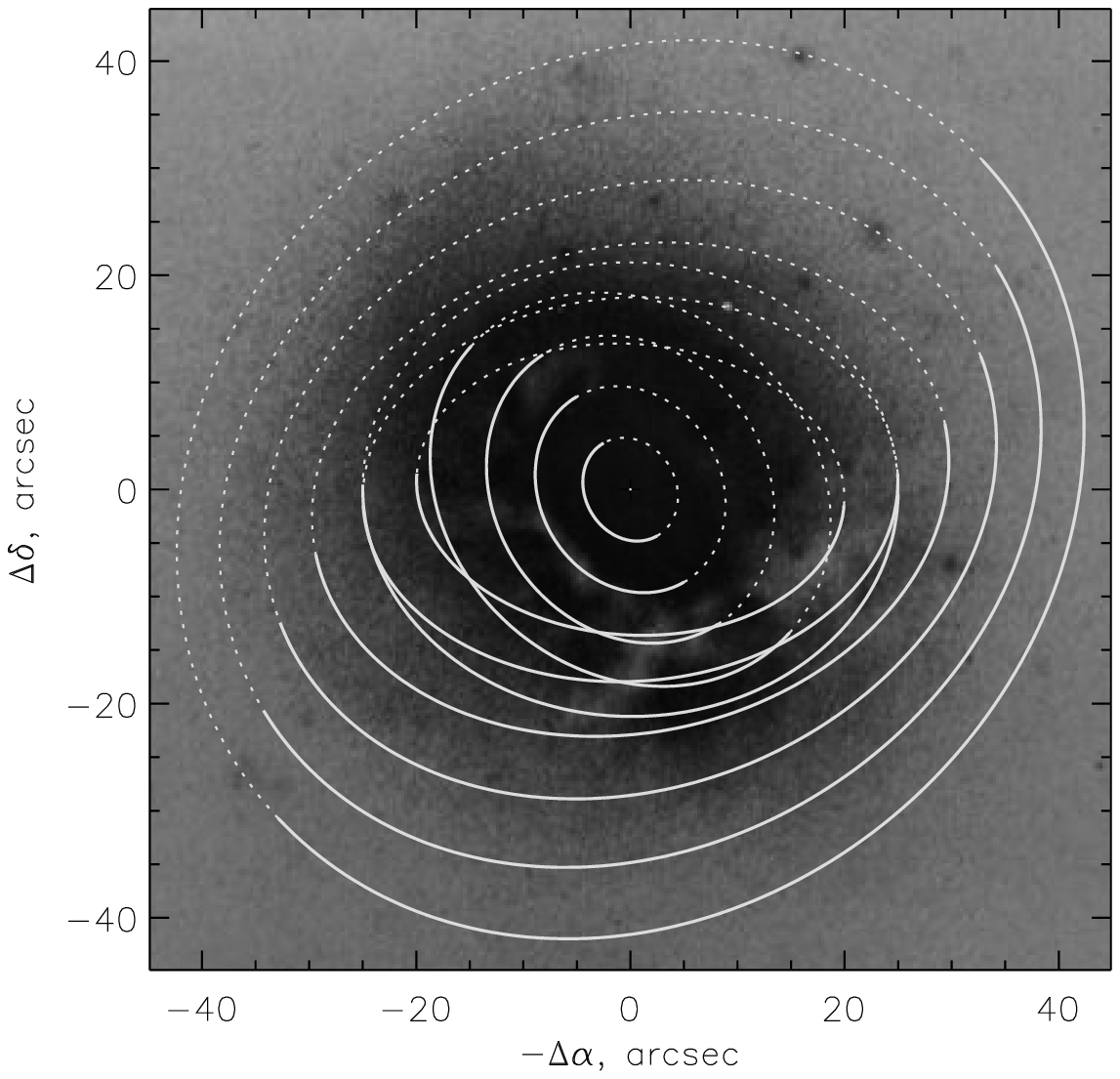}}
\caption{H$\alpha$ image of the regions located in the warped
outer disk of the galaxy (left panel) and the red-plate photograph
of the galaxy from Arp's (\cite{arp:Moiseev_n}) atlas (right panel).
The ellipses indicate the projections of circular orbits  (with a
radial step of $5''$). The dashed lines indicate the parts of the
orbits located behind the sky plane. Both the orbits in the inner
disk  (in the plane of the galaxy) and those in the outer warped
disk (the ring) are shown.}
  \label{fig_orbits:Moiseev_n}
\end{figure*}

The facts that radial velocities of the outer disk agree well with
the circular-model approximation (which takes into account the
warp of the plane of rotation) and the systemic velocity coincides
with the systemic velocity for the inner disk, indicates that we
observe steady-state gas rotation. This would not be the case of
the emission regions mentioned above belonged to a tidal tail as
suggested by Cair\'{o}s~et~al. (\cite{cairos2:Moiseev_n}). In
Fig.~\ref{fig_orbits:Moiseev_n} the sky-plane projections of
orbits both for the inner disk ($r<25''$) in the plane of the
galaxy, and for outer regions \mbox{($r$=20--$50''$)} are
presented. Both types of subsystems---the inner flat disk and the
outer warped disk---coexist in the  distance
interval $r$=20--$30''$ (2.0--3.5~kpc) and hence their orbits must
intersect. The collisions of gaseous clouds are possible
in the region of intersections. It is immediately apparent from
the right-hand panel of Fig.~\ref{fig_orbits:Moiseev_n} that the
powerful dust ring on the image of the galaxy is located inside
these radii. The dust ring must
correspond to the region where the gaseous disks collide. In this
region increase of gas density behind shock fronts results in the
formation of molecular and dust clouds like it happens in the
Galactic spiral shock  waves. An analysis of the ionization state of the
inner regions of the galaxy also points to a possible contribution
of shocks into the ionization of gas at $r\geq15''$, where the
dust ring starts Garc\'{i}a-Lorenzo  (\cite{cairos3:Moiseev_n}).

Perhaps it would be more correct to view the process not as direct collision, but as the ``fall'' of gas from the outer warped disk onto the main disk of the galaxy, because the angle between the two disks decreases
toward the center, i.e., precessing orbits of gaseous clouds in the outer disk approach the stellar disk of the galaxy and clouds fall onto this stellar disk.

When referring to the gaseous subsystem to which the outer
emission regions belong we usually employed the term ``the warped
disk''. However, strictly speaking, we are dealing with a
``ring'', because at  $r<2$~kpc only the gas that rotates in the
plane of the galaxy is observed. Thus outer HII regions are
located inside a broad and thin gaseous ring whose tilt with
respect to the main disk increases with radius and reaches
$50\degr$ at  $r=50''$ (5.8~kpc), which
corresponds to the $R_{25}$ optical radius. This fact confirms the
hypothesis of Whitmore~et~al. (\cite{w90:Moiseev_n}) that Arp~212 is
an object ``related to polar-ring galaxies''. Arp~212 can be
safely put into this class of galaxies given the fact that rings
in PRG are not always strictly polar---warped tilted rings are
also commonly found. There is a well-known example---the NGC\,660
galaxy, where the outer ring is tilted with respect to the main
disk by about $60\degr$ (Arnaboldi  \&  Galletta, \cite{arnaboldi93:Moiseev_n}). However,
unlike classical PRG, the polar ring in Arp~212 cannot be seen on
optical images (except for the HII regions mentioned above) and
shows up mostly in neutral hydrogen.

\begin{figure*}[tbp]
\centerline{\includegraphics[width=15 cm]{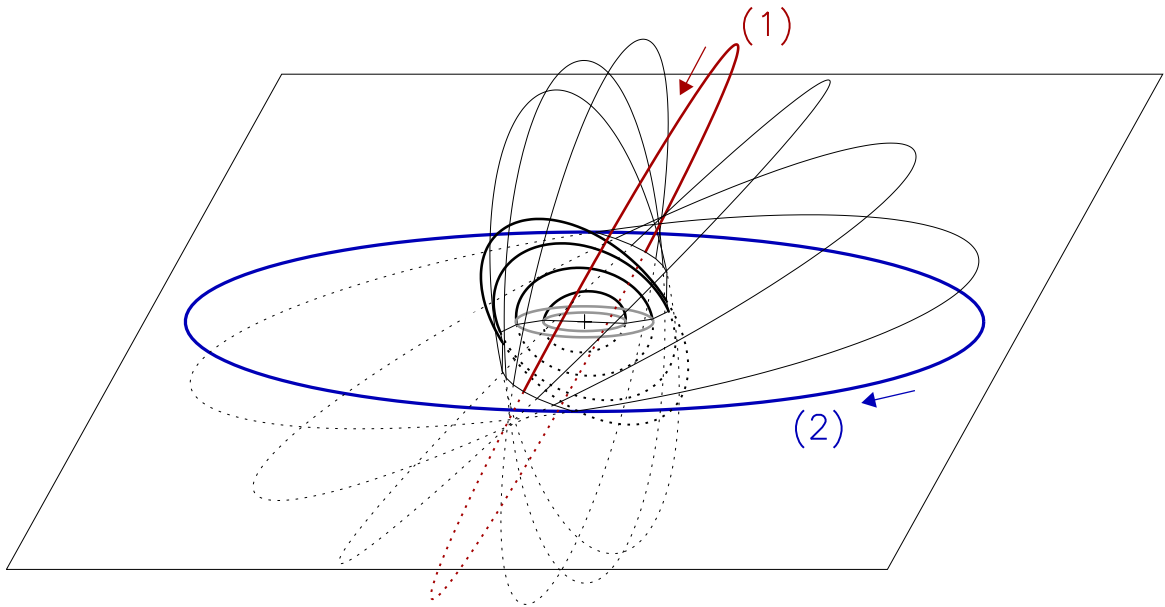}}
\caption{Orientation of circular orbits in the warped gaseous ring
with respect to the main plane of the galaxy. The solid black
lines indicate the orbits computed in terms of the two-dimensional
model of the velocity field. The thin black lines represent the
possible warp of the disk at large galacticentric distances. The
gray ellipses are the orbits of the inner disk of ionized gas in
the plane of the galaxy. The red and blue ellipses are  the two possible orientations of the outer regions of the HI disk ((1) and (2)). The arrows indicate the direction of rotation. The dashed lines correspond  to the parts of orbits that lie below the main plane. Inner regions of the main
disk rotate in the counterclockwise direction.}
\label{fig_orbits2:Moiseev_n}
\end{figure*}

In Fig.~\ref{fig_orbits2:Moiseev_n} we represent the space
location of the warped ring with respect to the plane of the
galaxy. We constructed the kinematical model solely for ionized
gas, although it is evident that we are dealing with a more global
gaseous disk interspersed with isolated regions of ongoing star
formation. They trace the kinematics of the neutral-hydrogen disk
with sufficiently high surface density inside the optical
radius (Li et al., \cite{li:Moiseev_n}). Observations at 21~cm
(Li et al., \cite{li:Moiseev_n}) indicate that we see a single gaseous
structure extending from the center to the outermost regions at
\mbox{$r\approx$ 20--22~kpc}. Unfortunately, the low angular
resolution of HI data prevents a detailed analysis of gas
motions outside the optical disk. The only more or less
definitive conclusion concerning the kinematics of gas in outer
regions  based on the data of Li~et~al. (\cite{li:Moiseev_n})  is that at $r\geq15$~kpc gas rotates about an axis, close to the rotation axis of the galaxy, but the direction of gas rotation is opposite
to the rotation of inner regions.

Let us try to extrapolate the behavior of the gaseous disk beyond
the optical radius. It is safe to assume that the outer disk (the
ring) continues to warp at $r>50''$. In this case its plane should
become polar to the stellar disk  of the galaxy not too far from
the center (position (1) in Fig.~\ref{fig_orbits2:Moiseev_n}). The
continued warp of the gaseous ring at these galactocentric
distances is not just an extrapolation of our kinematical model,
but is also consistent with the available observational data for
PRG with strongly warped disks, e.g., NGC\,2655
(Sparke et al., \cite{sparke08:Moiseev_n}). Such a distortion of the shape of the polar ring is believed to be due to the precession of orbits in a
nonspherical gravitational potential. In case of Arp~212, if the
warp continues in the same direction, the angle between the
orbital plane and the plane of the galaxy should decrease with
further increase of $r$, so that the polar ring ``tips over'' onto
the plane of the galaxy (position (2) in
Fig.~\ref{fig_orbits2:Moiseev_n}). It is also evident from this
figure that in case of such a ``tipover'' the direction of
rotation of outer regions should be opposite that of the inner
parts of the disk, in accordance with HI observations.

However, there are reasons to doubt the reality of the scenario
(2). If the outer gas is located close to the plane of the galaxy
then the direction of the orbital momentum should be reversed as a
result of precession. It is a very peculiar pattern, because
precession is usually a result of withdrawal or redistribution of
the angular momentum, whereas in our case we have to invent a
mechanism that would transfer substantial angular momentum to
gaseous clouds in order to reverse the direction of rotation. It
must be a very specific mechanism, because such complex warps of
disks (rings) are not commonly observed in PRG. Of course, we can
assume that in Arp~212 we observe two independent systems of
neutral hydrogen---one of them has produced the polar ring and the
other, the counter-rotating outer disk. However, we consider it
very unlikely to have caught the galaxy during the stage when it
interacts simultaneously with two different gaseous subsystems.
Moreover, the available HI observations are rather indicative of
the existence of a single structure in the distribution of neutral
hydrogen.

\begin{figure*}[tbp]
\centerline{\includegraphics[width=16 cm]{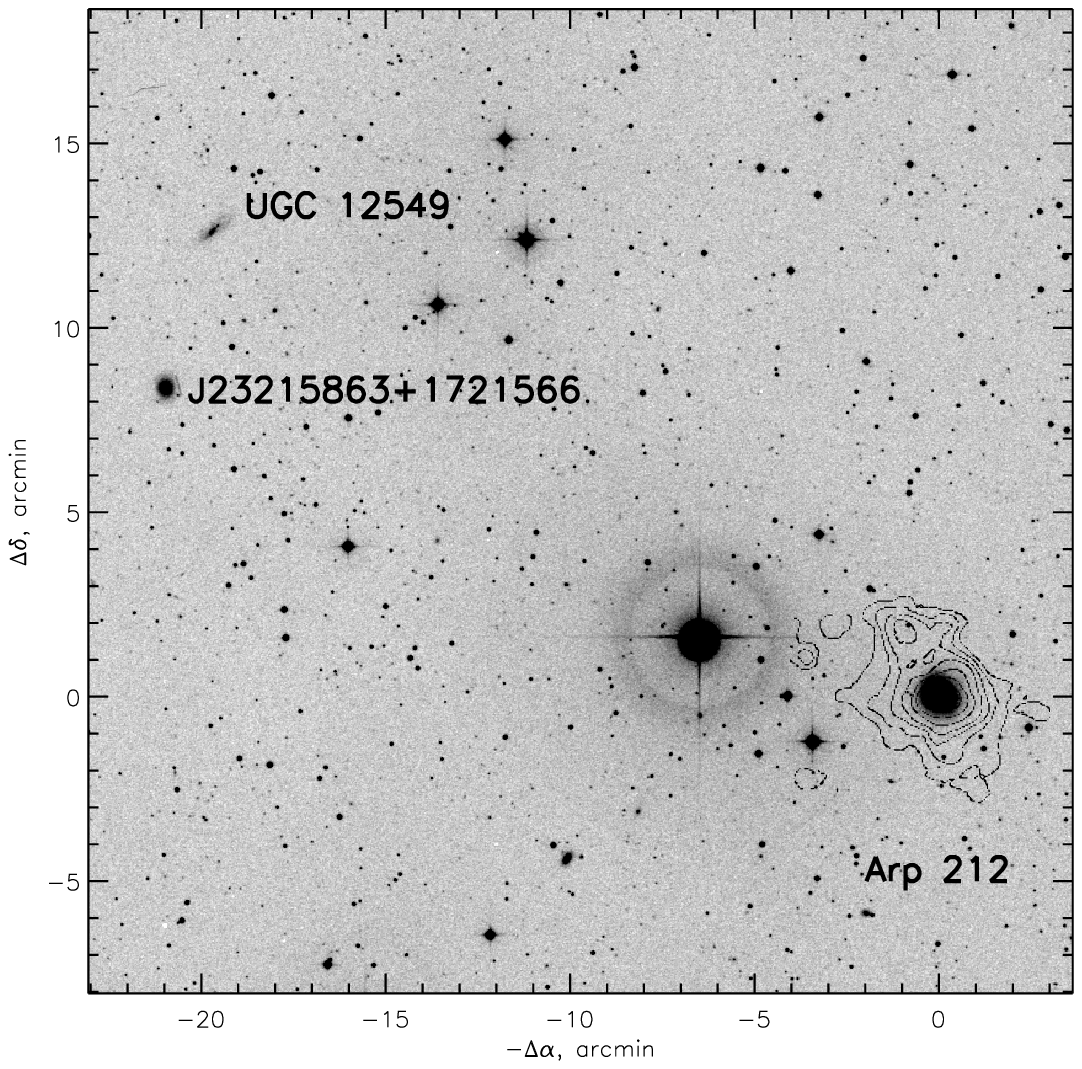}}
\caption{POSS2 field in the vicinity of  Arp~212 with HI density
contours (Li et al., \cite{li:Moiseev_n}) superimposed. Two possible companions
are indicated.} \label{fig_dss:Moiseev_n}
\end{figure*}

We therefore consider scenario (1), where outer parts of the HI
ring are located in the vicinity of the polar plane, to be more
realistic, and, moreover, it is also consistent with the results
of observations. Indeed, according to
Li~et~al. (\cite{li:Moiseev_n}), outer HI density contours are
appreciably extended along  $PA_2\approx25\degr$ and their axial
ratio is about 2/3 (see also Fig.~\ref{fig_dss:Moiseev_n}), which
corresponds to  $i_2\approx50\degr$ in case of thin-disk
projection. We further assume that the Western part of the HI disk
is the closest to us and recall that velocities in its Northern
half are redshifted, to find from formula (\ref{eq1:Moiseev_n})
that outer parts of the HI ring  are located at angle
$\delta=82\degr$, and this result agrees well with the polar-disk
hypothesis. The weak point of our hypothesis  consists in the lack
of accurately known kinematical parameters for the gaseous disk at
\mbox{$r>50''$.} Further high-resolution HI data are needed.

Such a behavior of the warped polar ring can be explained by
assuming that the distribution of gravitational potential in the
galaxy is not spherical. In this case the orbits of the polar
component should precess. One of the most famous PRG---NGC\,2685,
where the HI kinematics can be explained by the deviation of the
distribution of potential from spherical
symmetry (J\'{o}zsa et al., \cite{jozsa04:Moiseev_n})---provides a beautiful example
of a strong warp of the polar ring. The researchers have
repeatedly discussed the stability of gaseous disks in the
gravitational potential of three-axial galaxies.  In this case
stable are disks that lie in the planes perpendicular to the major
or minor axes of the system. The example of the gaseous disk in
the elliptical galaxy Cen~A analyzed by
van~Albada~et~al. (\cite{vanalbada:Moiseev_n}) somewhat resembles
the warped disk in Arp\,212: in the inner region the gaseous disk
is orthogonal to the stellar body of the galaxy, but at large
radii the gaseous disk in Cen~A warps so that
its outer parts tend to be located in the rotation plane of the
stellar subsystem and rotate in the opposite direction. However,
in   Arp~212 the situation is quite the opposite --   the outer
parts of the HI disk are located in the polar plane and with
decreasing radius orbits precess and approach the plane of the
stellar disk. In any case, the hypothesis that outer parts of the
polar ring are located in one of the principal planes of the
three-axial halo (the contribution of the disk to the potential is
small at $r=2R_{25}$) appears reasonable, although we bear in mind
that the available data are so far insufficient for a definitive
conclusion about the three-axial distribution of the potential.
Note also that the suggestion for oblate distribution of the
potential can be found inside the optical radius. Indeed, it
follows from Fig.~\ref{fig_rc:Moiseev_n}  the velocities of
rotation of outer HII regions in the inclined-disk model are
somewhat higher than in the inner disk. This is not a surprise,
because at these radii an oblate stellar disk must contribute appreciably to the total gravitational potential.

\subsection{The Origin of the Outer Gaseous Subsystem}

According to modern scenarios, the formation of polar rings is due
to the interaction between galaxies: their mergers or accretion of
the matter of a companion galaxy onto the main
galaxy (Bournaud   \& Combes, \cite{bc03:Moiseev_n}; Combes, \cite{comb06:Moiseev_n}). In this case either
the companion galaxy or tidal features remaining after its
disruption should be found in the neighborhood of the galaxy
considered. Above we already pointed out some similarities between
Arp~212 and NGC 2655---another galaxy with a warped polar disk.
Deep images of this galaxy, presented by
Sparke~et~al. (\cite{sparke08:Moiseev_n}), show low surface
brightness tidal features within two optical diameters. However,
our image of Arp~212 lacks tidal features brighter than
$25.2^{m}/\Box''$~(see Section~\ref{obs:Moiseev_n}).

\begin{figure}[tbp]
\centerline{\includegraphics[width=8. cm]{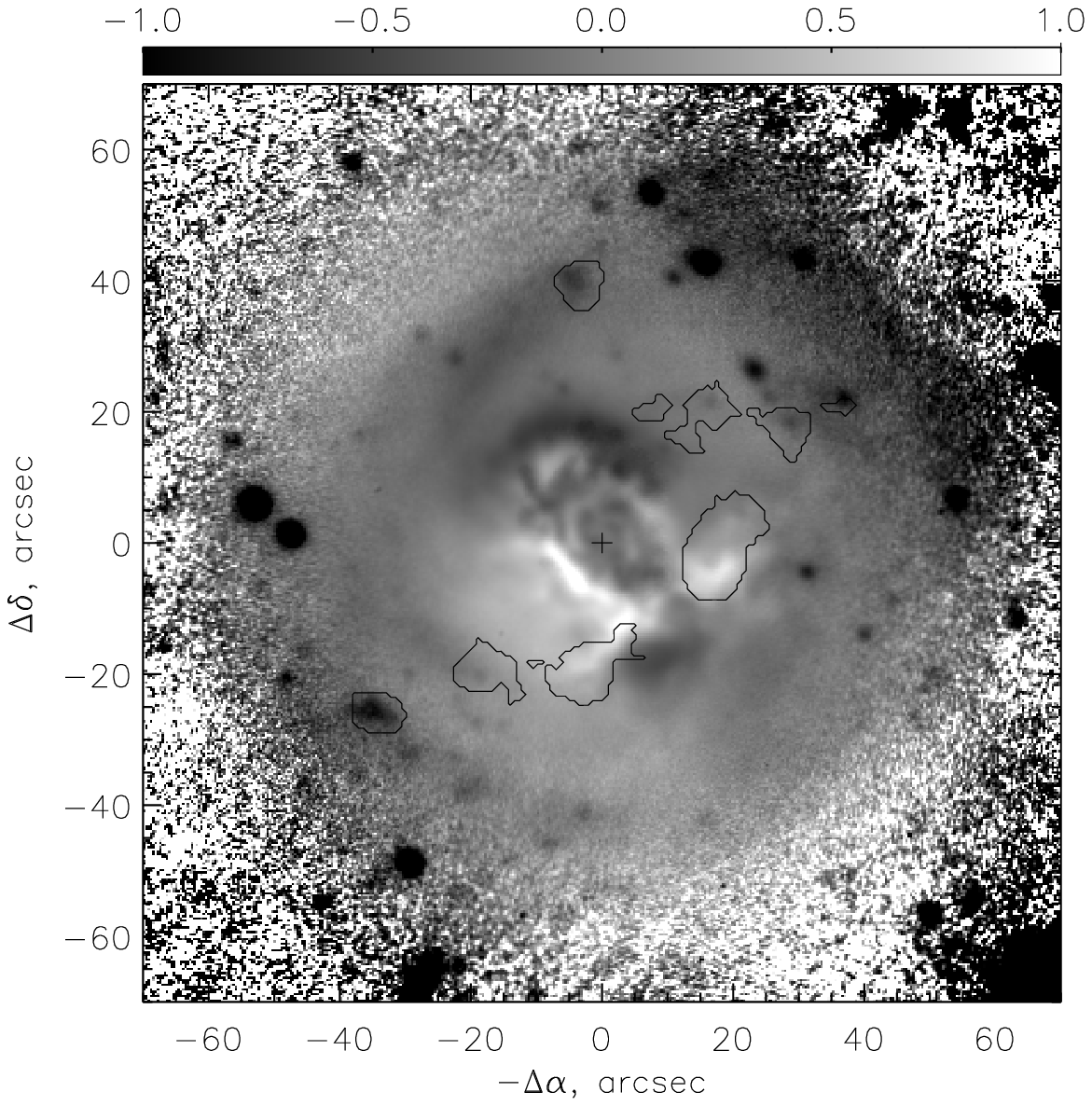}}
\caption{Residual brightness in the magnitude scale (observations
minus the bulge and disk model). The contours indicate the
locations of the outer HII regions that belong to the polar ring.}
\label{fig_image2:Moiseev_n}
\end{figure}

In Fig.~\ref{fig_image2:Moiseev_n} one can see the
surface-brightness distribution on the R$_c$ image of the galaxy
after the subtraction of the two-dimensional model consisting of a
bulge and exponential disk (see Section~\ref{obs:Moiseev_n}). The
distribution of residual surface brightness is mostly due to the
star-forming regions and dust lanes in the disk. We failed to find
any stellar tidal features associated with a disrupt companion
inside the optical radius. The only exception is the  ``filament''
\mbox{$r$=20--$40''$} North of the nucleus, where one of the outer
HII regions projects onto. However, this may be  a fragment of the
spiral pattern in the stellar disk. The bright dust lane located
to the South and Southwest of the nucleus, which coincides with
three outer HII regions, is the only structural feature on this
image that is undoubtedly associated with the warped polar ring.
Here we must be observing dust inside the polar ring. This
hypothesis is consistent with our previous conclusion that the
Southern half of the ring is the nearest to the observer. This
dust lane, like the HII regions, is not indicative of the remnants
of a companion, but rather of ongoing star formation inside the
polar ring.

The lack of stellar remnants of a dusrupted companion may imply that the polar ring possibly formed as a result of the capture of gas from a companion galaxy. Such a process of the formation of polar rings has been rather
thoroughly studied using various methods including numerical simulations
(Reshetnikov  \& Sotnikova, \cite{reshetnikov97:Moiseev_n}; Bournaud   \& Combes, \cite{bc03:Moiseev_n}).

According to Bournaud   \& Combes (\cite{bc03:Moiseev_n}), most of the polar rings form in accordance with the ``accretion'' scenario. The NED database
contains only one galaxy---UGC~12549, inside the region of
$1\degr$ radius (414~kpc) centered on Arp~212 within
\mbox{$\pm1000\km$} interval of the line-of-sight velocity of
Arp~212. It is an Im-type galaxy, which has almost exactly the
same systemic velocity as Arp~212 --- $1634 \km$, but is $3^m$
fainter in terms of integrated magnitude. Its distance from
Arp~212 (projected onto the sky-plane) is 23\farcm6 (163 kpc), or
$15D_{25}$. This galaxy is very rich in gas: the total 21-cm flux
of UGC~12549 is only twice lower than that of Arp~212
(Springob et al., \cite{springob05:Moiseev_n}). The total HI mass in UGC~12549 then
must be $10^9 M_{\odot}$, and the HI-to-the-total-mass ratio in
this galaxy must be several times higher than in Arp~212, where it
is about 10\%. In Fig.~\ref{fig_dss:Moiseev_n} we show both
galaxies according to the Palomar DSS2. There is a
brighter galaxy, 2MASX~J23215863+1721566, with unknown redshift
4\farcm5 (31\,kpc) South of UGC~12549. These two galaxies may form
a pair. It is evident from Fig.~\ref{fig_dss:Moiseev_n} that HI
density contours (i.e., the projection of outer regions of the
polar ring) in Arp~212 are extended in the direction of this pair
of galaxies. There may be a faint tidal tail between the galaxies.
However, so far we have no data about the distribution of neutral
hydrogen beyond \mbox{20-30\,kpc} from the center of  Arp~212. New
observations are required in order to confirm or disprove the
hypothesis that UGC~12549 served as a donor for the formation of
the polar ring in Arp~212.

In conclusion we address the problem of the burst of star
formation in Arp\,212. According to photometric data, all
star-forming regions in the galaxy are $6.1\pm0.6$~Myr old
(Mart\'{i}nez-Delgado, \cite{delgado:Moiseev_n}), i.e., the most recent burst of star
formation started almost simultaneously throughout the entire
galaxy. The dynamical time scale (orbital period) for outer
emission filaments is 0.2--0.3~Gyr, and that of the outer regions
of the HI disk, 0.5--1~Gyr. It is safe to assume that the outer
gaseous subsystem---after it was captured by the main
galaxy---formed the polar disk whose gas began, as a result of
orbital precession, to  ``fall'' onto the plane of the stellar
disk of the galaxy. After sufficient amount of gas was
accumulated, a burst of star formation began. The fact that the
age of the burst of star formation is smaller that the dynamical
age of the gaseous disk may imply that such bursts may have
occurred more than once in the galaxy considered.

\section{Conclusion}

\label{results:Moiseev_n}

We constructed and analyzed the velocity field of ionized gas in
the peculiar galaxy Arp\,212 and found it to have two
kinematically distinct subsystems of rotation gas---the inner disk
and outer HII regions. The inner disk is located within 3.5 kpc of
the center and ionized gas in this disk rotates in the plane
coincident with the plane of the stellar disk. Note that the inner
part of the ionized-gas disk is exactly coincident with the
earlier known molecular-gas disk. The second subsystem of ionized
gas is located at a galactocentric distance of
\mbox{$r$=2--6\,kpc} and consists of isolated HII regions whose
orbits are tilted by significant angles to the stellar disk. Our
own and published data on the kinematics of molecular, ionized,
and neutral gas can be explained in terms of the following model.
Most of the HI mass in the outer parts of the galaxy concentrates
in a broad ring with radius of about 20\,kpc. Outer ring regions
rotate in the plane that is orthogonal to the plane of the stellar
disk. With decreasing galactocentric distance the orbits of
gaseous clouds precess and approach the plane of the disk. This
precession is due to the nonspherical   (maybe a triaxial) distribution of the gravitational potential in the galaxy.
A burst of star formation takes place in the inner regions of the
warped polar ring. The angle between the ring and the plane of the
galaxy is equal to $50\degr$ at $r=R_{25}$, it continues to
decrease with decreasing  radius, and at $r\approx$2--3~kpc the gas from the ring falls onto the plane of the galaxy.   The powerful dust ring in the central region of the galaxy is indicative of the fall of gas from the polar ring onto the plane of the stellar disk of  Arp\,212. The polar ring may have formed as a result of the interaction with the gas-rich dwarf galaxy UGC~12549, however, new observations of the distribution and kinematics of neutral hydrogen are required to either confirm or disprove this scenario.

\begin{acknowledgements}
This work is based on observations collected with the 6-m telescope of the Special Astrophysical Observatory (SAO) of the Russian Academy of Sciences
(RAS), operated under the financial support of the Science Department of Russia (registration number 01-43). This research has made use of the NASA/IPAC Extragalactic Database (NED) which is operated by the Jet Propulsion Laboratory, California Institute of Technology, under contract with the
National Aeronautics and Space Administration. I am grateful to Vladimir Resehtnikov for a number of valuable comments that he made after reading the draft of the paper. This work was supported by the Russian Foundation for Basic Research (project no. 06-02-16825) and grant no. MK1310.2007.2 of the President of the Russian Federation.
\end{acknowledgements}


\begin{thebibliography}{}

\bibitem[2005]{afanas05:Moiseev_n}
Afanasiev V.L.,  Moiseev A.V., 2005, Astronomy Letters, {\bf 31}, 193; astro-ph/0502095


\bibitem[1993]{arnaboldi93:Moiseev_n}
Arnaboldi M., \&  and Galletta G., 1993, \aap, 268, \textbf{411}

\bibitem[1966]{arp:Moiseev_n}
Arp H., 1966, \apjs, \textbf{14}, 1

\bibitem[1989]{beg89:Moiseev_n}
Begeman K.G., 1989, \aap, \textbf{223}, 47

\bibitem[2003]{bc03:Moiseev_n}
Bournaud F., \& Combes F., 2003, \aap, \textbf{401}, 817

\bibitem[1991]{br:Moiseev_n}
Brosch N., \& Loinger F., 1991,  \aap, \textbf{249}, 327

\bibitem[2001a]{cairos1:Moiseev_n}
Cair\'{o}s L.M., V\'{i}lchez J.M., G\'{o}nzalez-P\'{e}rez J.N., et al., 2001a, \apjs, \textbf{133}, 321

\bibitem[2001b]{cairos2:Moiseev_n}
Cair\'{o}s L.M.,   Caon B., V\'{i}lchez J.M., et al., 2001b, \apjs,  \textbf{133}, 321

\bibitem[2007]{coccato07:Moiseev_n}
Coccato L., Corsini E.M., Pizzella A., \&  Bertola F., 2007, \aap, \textbf{465}, 777

\bibitem[1997]{courteau97:Moiseev_n}
Courteau S., 1997, \aj, \textbf{114}, 2402 ( 

\bibitem[2006]{comb06:Moiseev_n}
Combes F., 2006, in \textit{Mass Profiles and Shapes of Cosmological Structures}, Eds.: Mamon G.A., Combes F., Deffayet C., Fort B., (EAS Publications Series, \textbf{20}, p. 97; astro-ph/0508434

\bibitem[1969]{demoulin:Moiseev_n}
Demoulin M.-H.,  1969, \apj, \textbf{157}, 69

\bibitem[2008]{cairos3:Moiseev_n}
Garc\'{i}a-Lorenzo B., Cair\'{o}s L.M., Caon N. et al., 2008, \apj, {\bf 677}, 201

\bibitem[2004]{jozsa04:Moiseev_n}
J\'{o}zsa G., Oosterloo T., \& Klein U., 2004, in {\it Proceedings of ``Baryons in Dark Matter Halos'', Novigrad, Croatia,  2004}, Eds.:\ Dettmar R., Klein U., Salucci P.,  (Published by SISSA, Proceedings of Science), p. 108

\bibitem[1993]{li:Moiseev_n}
Li J.G., Seaquist E.R., Wrobel J.M. et al., 1993,  \apj, \textbf{413}, 150

\bibitem[2007]{delgado:Moiseev_n}
Mart\'{i}nez-Delgado I., Tenorio-Tagle  G.,  Mu\~{n}oz-Tu\~{n}\'{o}n C., et al., 2007, \aj, \textbf{133}, 2892

\bibitem[2002]{mois02:Moiseev_n}
Moiseev A.V., 2002, Bull. SAO  {\bf 54}, 74; astro-ph/0211104

\bibitem[2008]{mois08:Moiseev_n}
Moiseev A.V., \& Egorov O., Astrophysical Bulletin, 2008, \textbf{63}, 193; arXiv:0805.2367 [astro-ph]

\bibitem[2000]{mois00:Moiseev_n}
Moiseev A.V., \& Mustsevoi V.V., 2000, Astronomy Letters, {\bf 26}, 565; astro-ph/0011225

\bibitem[2004]{mois04:Moiseev_n}
 Moiseev A.V., Vald{\' e}s J.R. \& Chavushyan V.H., 2004, \aap, \textbf{421}, 433

\bibitem[1993]{reshetnikov93:Moiseev_n}
Reshetnikov V.P., Hagen-Thorn V.A. \&  Yakovleva V.A., 1993, \aap, {\bf 278}, 351

\bibitem[1997]{reshetnikov97:Moiseev_n}
Reshetnikov V., \& Sotnikova N., 1997, \aap, \textbf{325}, 933

\bibitem[2007]{labuda07:Moiseev_n}
Shalyapina L.V., Merkulova O.A., Yakovleva V.A., \&
Volkov E.V., 2007, Astronomy Letters, 33, 520,  \textbf{33}, 585 

\bibitem[2004]{silafan04:Moiseev_n}
Sil'chenko O.K., \& Afanasiev V.L., 2004, \aj, \textbf{127}, 2641

\bibitem[2008]{sparke08:Moiseev_n}
Sparke L.S., van Moorsel G., Erwin  P., \& Wehner E.M.H., 2008, \aj, \textbf{135}, 99

\bibitem[2005]{springob05:Moiseev_n}
Springob C.M., Haynes M.P., Giovanelli R., \& Kent  B.R., 2005, \apjs, \textbf{169}, 149

\bibitem[1981]{thuan:Moiseev_n}
Thuan T.X., \& Martin  G.E., 1981, \apj, \textbf{247}, 823

\bibitem[1982]{vanalbada:Moiseev_n}
van Albada T.S., Kotanyi C.G., \&  Schwarzschild M., 1982, \mnras, \textbf{198}, 303

\bibitem[1990]{w90:Moiseev_n}
Whitmore B.C., Lucas R.A., McElroyet D.B. et al., 1990, \aj, \textbf{100}, 1489
\bibitem[1992]{co:Moiseev_n}
Yasuda N., Fujisawa K., Sofue Y. et al., 1992, \pasj, \textbf{44}, 1


\end{thebibliography}
\end{document}